\pdfoutput=1
\documentclass[twocolumn,floatfix,tighten,times]{aastex631}

\usepackage{graphicx}    

\usepackage{float} 

\usepackage{dcolumn}     
\usepackage{bm}          

\usepackage{amsmath} 
\usepackage{amssymb} 
\usepackage{amsfonts} 
\usepackage{latexsym}
\usepackage{enumitem} 

\usepackage{booktabs} 

\usepackage{natbib}

\usepackage[normalem]{ulem}

\usepackage[english]{babel}
\usepackage[expansion=all,babel]{microtype}

\usepackage{cleveref}
\usepackage{mathtools}

\usepackage{multirow}

\usepackage{rotating}

\newcommand{\sat}[1]{{#1}_{\mathrm{sat}}}
\newcommand{\sym}[1]{{#1}_{\mathrm{sym}}}
\newcommand{\eff}[1]{{#1}_{\mathrm{eff}}}

\newcommand{\meffnSNM}{m_{\mathrm{eff;}n}^{\mathrm{SNM}}}
\newcommand{\meffnPNM}{m_{\mathrm{eff;}n}^{\mathrm{PNM}}}

\newcommand{\tth}[1]{{#1}_{\mathrm{th}}}
\newcommand{\be}{\begin{equation}}
\newcommand{\ee}{\end{equation}}
\newcommand{\bea}{\begin{eqnarray}}
\newcommand{\eea}{\end{eqnarray}}

\usepackage{txfonts}

\newcommand{\Msun}{\mathrm{M}_\odot}

\makeatletter
	\newcommand{\vast}{\bBigg@{2.85}}
\makeatother

\shorttitle{General purpose EOSs for astrophysical simulations}
\shortauthors{Raduta \& Beznogov}

\begin{document}

\title{New ab initio constrained extended Skyrme equations of state for simulations of neutron stars, supernovae and binary mergers:
II. Thermal response in the suprasaturation density domain}

\author[0000-0001-8421-2040]{Adriana R. Raduta}
\email{araduta@nipne.ro}
\affiliation{National Institute for Physics and Nuclear Engineering (IFIN-HH), RO-077125 Bucharest, Romania}

\author[0000-0002-7326-7270]{Mikhail V. Beznogov}
\email{mikhail.beznogov@nipne.ro}
\affiliation{National Institute for Physics and Nuclear Engineering (IFIN-HH), RO-077125 Bucharest, Romania}

\date{\today}

\begin{abstract}
Numerical simulations of core-collapse supernovae, mergers of binary neutron stars and formation of stellar black holes, which employed standard Skyrme interactions, established clear correlations between the evolution of these processes, characteristics of the hot compact objects, as well as neutrino and gravitational wave signals, and the value of effective nucleon mass at the saturation density. Unfortunately, the density dependence
of the effective mass of nucleons in these models does not align with the predictions of ab initio models with three body forces.
In this work, we investigate the thermal response for a set of extended Skyrme interactions that feature widely different density dependencies of the effective mass of the nucleons.
Thermal contributions to the energy density and pressure are studied along with a few thermal coefficients over wide domains of density, temperature and isospin asymmetry, relevant for the physics of hot compact objects.
For some of the effective interactions, the thermal pressure is negative at high densities.
This results in a situation where hot compact stars can support less mass before collapsing into a black hole compared to their cold counterparts.
Moreover, the higher the temperature, the lower the maximum mass that the hot star can support.
\end{abstract}


\section{Introduction}
\label{sec:intro}

The structure and composition of neutron stars (NSs) as well as the evolution of core-collapse supernovae (CCSNe) \citep{Janka_PhysRep_2007,Mezzacappa2015,Schneider_PRC_2017,Connor2018ApJ,Burrows2020MNRAS},
proto-neutron stars (PNSs) \citep{Pons_ApJ_1999,Pascal_MNRAS_2022},
binary neutron star (BNS) mergers \citep{Shibata_11,Rosswog_15,Baiotti_2017,Endrizzi_PRD_2018,Ruiz2020,Prakash_PRD_2021,Most_PRD_2023}
and the formation of black holes (BHs) in failed CCSNe \citep{Sumiyoshi_2007,Fischer_2009,OConnor_2011,hempel12}
depend upon the unknown properties of dense and strongly interacting baryonic matter.

The tremendous progress done by multimessenger astronomy of NSs over the past decade contributed unprecedented and valuable knowledge about the properties of states of matter that are impossible to produce and study in terrestrial laboratories. The high density behavior of the NS equation of state (EOS) was bracketed by measurements of pulsars with masses around or larger than $2~\Msun$ \citep{Demorest_Nature_2010, Antoniadis2013, Arzoumanian_ApJSS_2018, Cromartie2020, Fonseca_2021} and the interpretation of the outcome of the NS coalescence in the GW170817 event \citep{Abbott_PRL_2017} as a collapse of a hypermassive star into a BH, which can be translated into an upper limit
for the maximum mass that NS can sustain~\citep{Margalit_ApJ_2017,Rezzolla_ApJ_2018,Khadkikar_PRC_2021}.
The measurement of the combined tidal deformability of NSs with masses $1.17 \lesssim M/\Msun \lesssim 1.60$ in the GW170817 event~\citep{Abbott_PRL_2017,Abbott_PRX_2019} delivered the first ever constraint on the behavior of neutron-rich matter over the density range $1 \lesssim n/\sat{n} \lesssim 3$, where $\sat{n} \approx 0.16~\mathrm{fm}^{-3} \approx 2.7 \times 10^{14}~\mathrm{g/cm^3}$ represents the nuclear saturation density. Equatorial radii determination by {\em NICER} based on the analysis of the X-ray pulse profiles of millisecond pulsars with masses $1.4 \lesssim M/\Msun \lesssim 2.1$ \citep{Riley_2019,Miller_2019,Riley_may2021,Miller_may2021,Vinciguerra_ApJ_2024,Choudhury_ApJL_2024,Mauviard_2025} added further knowledge about the stiffness of neutron-rich EOS at suprasaturation domain.

A bulk of statistical inferences of NS EOSs were performed in the last few years, which systematically addressed the role of the EOS model, set of constraints, prior distributions, etc. For a discussion, see \citep{Beznogov_PRC_2023,Beznogov_ApJ_2024,Beznogov_PRC_2024} and references therein.
Despite this, the NS EOS remains largely unknown. Several factors contribute to this situation, including the not yet sufficiently well understood sensitivity of different astrophysical observations to various domains of density and isospin asymmetry, $\delta=(n_n-n_p)/n$, of nuclear matter (NM), a certain degeneracy with respect to the particle composition of dense matter, the unknown effective baryon-baryon interactions, and the still large error bars of astrophysical measurements; $n_n$ and $n_p$ represent the neutron and proton particle densities and $n=n_n+n_p$.

In the absence of constraints from nuclear physics experiments, the thermal response of dense matter, upon which the evolution of PNSs, CCSNe, BNS mergers and stellar BH formation depends, is even more mysterious. 
Numerical simulations of CCSNe~\citep{Schneider_PRC_2019b,Yasin_PRL_2020,Andersen_ApJ_2021}, BNS mergers~\citep{Fields_ApJL_2023,Raithel_PRD_2023}, and stellar BH formation~\citep{Schneider_ApJ_2020} have demonstrated that correlations exist between the evolution of these phenomena and the value of nucleon effective mass $\eff{m}$ at the saturation density. \cite{Schneider_PRC_2019b,Yasin_PRL_2020,Andersen_ApJ_2021} showed that large values of $\eff{m}$ favor high (low) values of the central density (temperature) in the cores of PNSs as well as lower PNS radii.
By playing on the PNSs' compactness, $\eff{m}$ also impacts the PNSs' oscillations and, consequently, the peak frequency of GWs~\citep{Andersen_ApJ_2021}. 
\cite{Schneider_PRC_2019b,Andersen_ApJ_2021} also proved that $\eff{m}$ influences the temperature, proton fraction ($Y_p=n_p/n$), density and radius of the neutrinosphere as well as neutrino energies and luminosities.
In failed CCSNe, the collapse into a BH happens earlier for EOSs with higher values of $\eff{m}$.
According to \cite{Fields_ApJL_2023}, $\eff{m}$ also imprints on the temperature and compactness of BNS mergers along with the strain and spectrum of GWs.
In all these circumstances, the sensitivity to $\eff{m}$ outsizes the sensitivity to any other parameter of the EOS,
including the stiffnesses of symmetric nuclear matter (SNM) and pure neutron matter (PNM).

The common feature of numerical simulations by \cite{Schneider_PRC_2019b,Yasin_PRL_2020,Schneider_ApJ_2020,Andersen_ApJ_2021,Fields_ApJL_2023} is the usage of
non-relativistic Skyrme interactions with monotonic behavior of $\eff{m}(n)$.
The huge advantage of Skyrme interactions is that analytical expressions are available for most thermodynamic and microscopic quantities~\citep{Constantinou_PRC_2014,Constantinou_PRC_2015}, which makes it possible to understand the role that the effective mass plays for the thermal pressure support or specific heat and, subsequently, in the evolution of various phenomena. 
While enlightening, these works do not guarantee that those correlations will persist if EOSs based on more realistic interactions are employed. 
In particular, ab initio calculations with three body forces predict that the nucleon effective mass as a function of density features an U-shaped behavior~\citep{Baldo_PRC_2014,Burgio_PRC_2020,Drischler_PRC_2021}. Depending on the forces and theoretical approaches that are used, for cold SNM the position of the minimum sits between $\sat{n}$~\citep{Drischler_PRC_2021} and $4\sat{n}$~\citep{Burgio_PRC_2020} and the value at the minimum is about $70\%$  of the bare mass. 
For PNM, the minimum sits around $2\sat{n}/3$ while the value at the minimum is about $88\%$~of the bare mass~\citep{Drischler_PRC_2021}.  
Complex density behaviors will obviously make it difficult to establish any connection between the evolution of phenomena in which wide ranges of densities are explored at different instances and spatial coordinates and the value of $\eff{m}(\sat{n})$. 
Still, the value that $\eff{m}$ takes at every single density will contribute to the fate of those phenomena, which means that understanding the thermal behavior of NM governed by more realistic forces is extremely important.

The aim of this paper is to systematically study the thermal response of a set of Brussels extended Skyrme interactions generated within a Bayesian inference of the EOS of dense matter~\citep{Beznogov_PRC_2024}. 
The advantage of employing Brussels extended Skyrme parametrizations consists, among others, in the possibility to qualitatively reproduce the non-trivial density dependence of the nucleon effective mass in ab initio models~\citep{Baldo_PRC_2014,Burgio_PRC_2020,Drischler_PRC_2021}. 
Usage of general purpose EOS tables based on such interactions in numerical simulations is expected to provide features distinct from those obtained when EOS tables based on more simple parametrizations are utilized.
In this way, we hope to contribute to a better understanding of the link between properties of NM and the evolution of CCSNe, PNSs, BNS mergers, and stellar BHs formation. In the longer term, our results may lead to constraints on the thermal behavior of NM.

The rest of the paper has the following structure. Sec.~\ref{sec:model} offers a brief review of the theoretical framework.
Two sets of effective interactions and five particular models that manifest extreme behaviors are discussed in Sec.~\ref{sec:NM}, where we also review key parameters of NM.
The five particular models correspond to RB(BBSk1), RB(BBSk2), RB(BBSk3), RB(BBSk4), and RB(BBSk5) general purpose EOS tables publicly available on \textsc{CompOSE} online repository~\citep{Typel_EPJA_2022}~\footnote{\url{https://compose.obspm.fr/}}; they were introduced in \citep{Raduta_AA_2025}, hereafter referred to as Paper I.
Selected properties of NSs built upon these models are presented in Sec.~\ref{sec:NS}. Sec.~\ref{sec:FiniteT} is dedicated to the thermal response. 
The stability of PNSs and remnants of BNS mergers is analyzed in Sec.~\ref{sec:PNS} in terms of the maximum gravitational and baryonic masses of isentropic stars. 
The conclusions are drawn in Sec.~\ref{sec:concl}.

In Sections 2, 3, and 5 the contributions of leptons and photons are disregarded.
Clusterization at densities lower than the saturation density and temperatures lower than the critical temperature of Coulomb instabilities, discussed in Paper I, is disregarded, too. 
More precisely, all results here correspond to homogeneous NM.

\section{Formalism}
\label{sec:model}

Hot strongly interacting NM is treated here within the self-consistent Hartree-Fock approach~\citep{Negele_PRC_1972,Vautherin1996} with Brussels extended Skyrme effective interactions~\citep{Chamel_PRC_2009}. 
In the absence of spin polarization and assuming zero electric charge, the energy density of bulk homogeneous matter is a sum of six terms,
\be
{\cal H}=k+h_0+h_3+\eff{h}+h_4+h_5.     
\label{eq:h}
\ee
Here, $k=\hbar^2 \tau/2m$ is the kinetic energy term and $2/m=1/m_{n}+1/m_{p}$, where $m_i$ with $i={n},{p}$ denotes the bare mass of nucleons; 
$h_0$ and $h_3$ are interaction terms that originate from the density-independent two-body term and the density-dependent term, respectively;
$\eff{h}$, $h_4$ and $h_5$ are the momentum-dependent terms of the interaction.
Each of the interaction terms can be expressed analytically in terms of particle number densities and kinetic energy densities, and the parameters of the effective interaction~\citep{Ducoin_NPA_2006,Beznogov_PRC_2024}:
\begin{align}
  &h_0 = C_0 n^2+D_0 n_3^2, \label{eq:h0} \\
  &h_3 = C_3 n^{\sigma+2}+D_3 n^{\sigma} n_3^2, \label{eq:h3} \\
  &\eff{h} = \eff{C} n \tau+\eff{D} n_3 \tau_3, \label{eq:heff} \\
  &h_4=\frac{t_4}{16} \left[ 3 n \tau -(2 x_4+1) n_3 \tau_3 \right] n^{\beta} \label{eq:h4} \\
  &h_5=\frac{t_5}{16} \left[ (4 x_5+5) n \tau + (2 x_5+1) n_3 \tau_3 \right] n^{\gamma} \label{eq:h5}.
\end{align}
In the equations above, $n=n_n+n_p$ and $n_3=n_n-n_p$ stand for the isoscalar and isovector particle number densities
and $\tau=\tau_{n}+\tau_{p}$ and $\tau_3=\tau_{n}-\tau_{p}$ denote the isoscalar and isovector kinetic energy densities.
$C_0$, $D_0$, $C_3$, $D_3$, $\eff{C}$, $\eff{D}$, $x_4$, $t_4$, $x_5$, $t_5$, $\sigma$, $\beta$ and $\gamma$ are
constants that define the effective interaction. Their values are determined upon fits of experimental nuclear data, results of ab initio models and astrophysical observations.

At finite temperature, $T$, the particle number densities
\be
n_i=\frac{g_i}{2 \pi^2} \int d k k^2 f_i(k),
\ee
and the kinetic energy densities
\be
\tau_i=\frac{g_i}{2 \pi^2} \int d k k^4 f_i(k),
\ee
are defined in terms of momentum distributions, $f_i(k)$, that describe the thermal population of momentum states
according to a Fermi-Dirac distribution,
\be
f_i(k)=\frac{1}{1+\exp\left[\left(\hbar^2 k^2/2 m_{\mathrm{eff;}i} +U_i - \mu_i \right)/T\right]}.
\ee
Here, $g_i=2$ is the spin degeneracy factor,
$\mu_i$ denotes the chemical potential of  the $i$-particle, $m_{\mathrm{eff;}i}$ stands for the effective mass of the $i$-particle,
\be
\frac{1}{m_{\mathrm{eff;}i}}=\frac{1}{m_i}+\frac{2}{\hbar^2} \left[\eff{\widetilde C}(n) n \pm \eff{\widetilde D}(n) n_3 \right],
\label{eq:meff}
\ee
where
\begin{align}
\begin{split}
    &\eff{\widetilde C}(n) = \eff{C} + \left[ 3 t_4 n^{\beta} +t_5 \left( 4 x_5+5\right) n^{\gamma}\right]/16, \\
    &\eff{\widetilde D}(n) = \eff{D} +\left[ -t_4 \left(2 x_4+1 \right) n^{\beta} +t_5  \left( 2 x_5+1\right) n^{\gamma}\right]/16.
  \end{split}
  \label{eq:CeffDeff_B}
\end{align}
$U_i=\partial {\cal H}/\partial n_i=U_{0_i}+U_{3_i}+U_{\mathrm{eff}_i}+U_{4_i}+U_{5_i}$ represents the single-particle potential of the $i$-particle.  

The locality of Skyrme interactions makes that nucleon effective masses are independent of temperature. In the limit of zero temperature, ${\eff{m}}_{;i}$ corresponds to the Landau effective mass defined in terms of density of single-particle states at the Fermi surface,
\be
\frac{1}{{\eff{m}}_{;i}}= \left . \frac{1}{\hbar^2 k_i} \frac{de_i}{dk_i} \right|_{k=k_{F;i}},
\ee
where $k_{F;i}$ stands for the Fermi momentum of the $i$-particle.
The expressions of the various terms that enter $U_i$ are the following,
\begin{align}
&U_{0_i} = 2 C_0 n \pm 2 D_0 n_3,
\label{eq:U0i} \\
&U_{3_i} = \left( \sigma+2 \right) C_3 n^{\sigma+1} + \sigma D_3 n^{\sigma-1} n_3^2 \pm 2 D_3 n^{\sigma} n_3,
\label{eq:U3i} \\
&U_{\mathrm{eff}_i} = \eff{C} \tau \pm \eff{D} \tau_3,
\label{eq:Ueffi} \\
\label{eq:U4i} \\
&U_{4_i} = \frac{t_4}8 n^{\beta-1} \left[ n \tau \left( 2 + x_4 + \frac{3 \beta}{2} \right)
-(1+2 x_4) \left( \frac{\beta}{2} n_3 \tau_3 + n \tau_i \right)
\right]\\
&U_{5_i} = \frac{t_5}{8} n^{\gamma-1} \Bigg[ n \tau \left( 2 + x_5 +\frac{5\gamma}{2} + 2 x_5 \gamma \right) + (1 + 2 x_5)  \nonumber \\ 
&\times  \left( \frac{\gamma}{2} n_3 \tau_3+ n \tau_i \right) \Bigg].
\label{eq:U5i}
\end{align}
In Eqs.~\eqref{eq:meff}, \eqref{eq:U0i}, \eqref{eq:U3i}, and \eqref{eq:Ueffi} the $\pm$ sign distinguishes the neutrons ($+$) from the protons ($-$).

The EOS stiffness is determined by pressure, 
\be
P=\frac23 k + h_0 +\left( \sigma+1\right)h_3 +\frac53 \eff{h}+ \left( \frac53 + \beta \right) h_4 + \left( \frac53 +\gamma \right) h_5.
\label{eq:p}
\ee
Considering that $\sigma$, $\beta$, $\gamma$ are positive, see Table III in \citep{Beznogov_PRC_2024}, it is clear that the EOS dependence of pressure outsizes the EOS dependence of energy.

Thermal contributions to state variables are conveniently gauged by taking the difference between the values that the quantity takes at finite and zero temperatures, $\tth{X}(n_n, n_p, T)=X(n_n, n_p, T)-X(n_n, n_p, T=0)$. In the case of energy density and pressure, one obtains~\citep{Constantinou_PRC_2014}
\be
\tth{e}=\sum_{i=n,p}\frac{\hbar^2}{2 m_{\mathrm{eff};i}} \left[ \tau_i(T)-\tau_i(T=0)\right]
\label{eq:eth}
\ee
and
\be
\tth{P}=\sum_{i=n,p} \frac{\hbar^2}{3 m_{\mathrm{eff};i}}
\left(1- \frac32 \frac{n}{m_{\mathrm{eff};i}} \frac{\partial m_{\mathrm{eff};i}}{\partial n}\right)\left[ \tau_i(T)-\tau_i(T=0)\right],
\label{eq:pth}
\ee
respectively.
Eqs.~\eqref{eq:eth} and \eqref{eq:pth} reveal that $\tth{e}$ and $\tth{P}$ have an explicit dependence on nucleons' effective masses.
Eq.~\eqref{eq:pth} shows that $\tth{P}$ also depends on the density dependence of nucleons' effective masses.
Notice, however, that via $\tau_i$ thermal energy density and thermal pressure depend on every term of the effective interaction.

In the low-temperature limit one can derive this dependence explicitly by means of the Sommerfeld expansion.
At the lowest order in temperature, it writes:
\begin{equation}
    \tau_i(T) - \tau_i(T=0) = \frac{T^2}{\hbar^4}\left(\frac{\pi}{3} \right)^{2/3} m_{\mathrm{eff};i}^2 n_i^{1/3}. 
\end{equation}
Substituting this expansion into Eqs.~\eqref{eq:eth} and \eqref{eq:pth}, one gets:
\begin{align}
    &\tth{e} \approx \frac{T^2}{2 \hbar^2} \left(\frac{\pi}{3} \right)^{2/3} \sum_{i=n,p} m_{\mathrm{eff};i} n_i^{1/3}, \\
    &\tth{P} \approx \frac{T^2}{3 \hbar^2} \left(\frac{\pi}{3} \right)^{2/3} \sum_{i=n,p} \left[ m_{\mathrm{eff};i}  - \frac{3}{2} \frac{\partial m_{\mathrm{eff};i}}{\partial n} n \right] n_i^{1/3}.
\end{align}
The expressions above are accurate for $T \lesssim 10-20$~MeV, depending on the effective interaction and the baryon number density.
One can see that in the low temperature limit $\tth{e}$ depends only on $\eff{m}$, while $\tth{P}$ depends only on $\eff{m}$ and its derivative with respect to $n$. 

\section{Nuclear matter}
\label{sec:NM}

\renewcommand{\arraystretch}{1.1}
\setlength{\tabcolsep}{3.9pt}
\begin{table}
\caption{Medians and 90\% CI of key properties of NM and NSs. The data on columns 3 and 4 refer to all models in the run~1 of \citep{Beznogov_PRC_2024} while data on columns 5 and 6 refer to the models selected according to the criteria in Sec.~\ref{sec:NM}.
For NM, provided are the saturation density ($\sat{n}$) of the SNM; the energy per nucleon ($\sat{E}$), compression modulus ($\sat{K}$),
skewness ($\sat{Q}$), and kurtosis ($\sat{Z}$) of the SNM at $\sat{n}$; 
the symmetry energy ($\sym{J}$), its slope ($\sym{L}$),
compressibility ($\sym{K}$), skewness ($\sym{Q}$), and kurtosis ($\sym{Z}$) at $\sat{n}$; the Landau effective mass of the nucleons
in the SNM ($m_{\mathrm{eff;\,n}}^\mathrm{SNM}$) and the Landau effective mass of the neutrons in the PNM ($m_{\mathrm{eff;\,n}}^\mathrm{PNM}$) at 0.16~fm$^{-3}$.
For NSs, provided are the maximum gravitational ($M_{\mathrm{G;TOV}}$) and baryonic ($M{_\mathrm{B;TOV}}$) masses;
the central particle density corresponding to the maximum mass configuration ($n_{\mathrm{c;TOV}}$);
the speed of sound squared ($c^{2}_{\mathrm{s;TOV}}$), energy density ($\rho_{\mathrm{c;TOV}}$) and pressure ($P_{\mathrm{c;TOV}}$) at $n_{\mathrm{c;TOV}}$;
radii ($R_{1.4}$, $R_{2.0}$) and tidal deformabilities ($\Lambda_{1.4}$, $\Lambda_{2.0}$) of NSs with masses equal to $1.4~\Msun$ and $2.0~\Msun$.}
\label{tab:Prop}
\centering
\begin{tabular}{lccccc}
\toprule
\toprule
	\multirow{2}{*}{Par.}               &  \multirow{2}{*}{Units}       & \multicolumn{2}{c}{all models (S1)}                  & \multicolumn{2}{c}{sel. models (S2)}   \\
\cmidrule(lr){3-4}
\cmidrule(lr){5-6}
                &                               & Med.     & 90\% CI              & Med.     & 90\% CI \\
\midrule
$n_\mathrm{sat}$ & $\mathrm{fm^{-3}}$             &    0.161 & $^{+0.0064}_{-0.0063 }$ &   0.161 &   $^{+0.0059}_{-0.0063}$ \\
$E_\mathrm{sat}$ & $\mathrm{MeV}$                &  -15.9   & $^{+0.33}_{-0.33}$      & -15.9 &     $^{+0.31}_{-0.34}$  \\
$K_\mathrm{sat}$ & $\mathrm{MeV}$                &  255     & $^{+34}_{-30}$         &  271  &     $^{+32}_{-16}$  \\
$Q_\mathrm{sat}$ & $\mathrm{MeV}$                & -383     & $^{+120}_{-90}$        & -331  &  $^{+100}_{-49}$\\
$Z_\mathrm{sat}$ & $\mathrm{MeV}$                & 1250     & $^{+750}_{-850}$       &  860  & $^{+360}_{-850}$\\
$J_\mathrm{sym}$ & $\mathrm{MeV}$                &   29.9   & $^{+1.7}_{-1.5}$       &   31  &    $^{+1.4}_{-1.2}$ \\
$L_\mathrm{sym}$ & $\mathrm{MeV}$                &   46.5   & $^{+12}_{-12}$         &  53.8 &      $^{+9.8}_{-8.5}$\\
$K_\mathrm{sym}$ & $\mathrm{MeV}$                & -122     & $^{+60}_{-46}$         & -130  &     $^{+28}_{-25}$\\
$Q_\mathrm{sym}$ & $\mathrm{MeV}$                &  590     & $^{+200}_{-220}$       &  386  &     $^{+76}_{-110}$ \\
$Z_\mathrm{sym}$ & $\mathrm{MeV}$                &  -2420   & $^{+830}_{-710}$       & -1850 &   $^{+530}_{-410}$ \\
$m_{\mathrm{eff;\,n}}^\mathrm{SNM}$ & $m_{\mathrm{n}}$ &  0.539  & $^{+0.2}_{-0.088}$      &  0.634 &    $^{+0.14}_{-0.085}$\\
$m_{\mathrm{eff;\,n}}^\mathrm{PNM}$ & $m_{\mathrm{n}}$ &  0.892  & $^{+0.087}_{-0.13}$     &  0.915 &    $^{+0.062}_{-0.091}$\\
$M_{\mathrm{G;TOV}}$    & $\Msun$                     & 2.13  & $^{+0.14}_{-0.088}$     &  2.09 &    $^{+0.11}_{-0.048}$\\
$M_{\mathrm{B;TOV}}$    & $\Msun$                     & 2.56  & $^{+0.19}_{-0.13}$     &   2.5 &    $^{+0.14}_{-0.065}$\\
$c^{2}_{\mathrm{s;TOV}}$  & $c^2$                      & 0.885  & $^{+0.11}_{-0.3}$     &  0.915 &     $^{+0.069}_{-0.25}$  \\
$n_{\mathrm{c;TOV}}$    & $\mathrm{fm}^{-3}$          & 1.1   & $^{+0.093}_{-0.11}$     &  1.13 &   $^{+0.063}_{-0.095}$\\
$\rho_{\mathrm{c;TOV}}$ & $10^{15}~\mathrm{g/cm^3}$   & 2.56  & $^{+0.26}_{-0.27}$      &  2.64 &    $^{+0.16}_{-0.23}$\\
$P_{\mathrm{c;TOV}}$    & $10^{36}~\mathrm{dyn/cm^2}$ & 1.17  & $^{+0.24}_{-0.3}$       &  1.21 &     $^{+0.11}_{-0.26}$\\
$R_{1.4}$          & $\mathrm{km}$              &  12.1  & $^{+0.53}_{-0.59}$         & 12.2 &    $^{+0.38}_{-0.33}$\\
$\Lambda_{1.4}$    & --                         &  372   & $^{+130}_{-100}$           & 392  &     $^{+91}_{-63}$ \\
$R_{2.0}$          & $\mathrm{km}$              & 11.4   & $^{+0.82}_{-0.65}$          &  11.3 &    $^{+0.67}_{-0.47}$  \\ 
$\Lambda_{2.0}$    & --                         & 21     & $^{+17}_{-8.5}$            &  19.2 &      $^{+13}_{-6}$ \\
\bottomrule
\bottomrule
\end{tabular}
\end{table}
\renewcommand{\arraystretch}{1.0}
\setlength{\tabcolsep}{2.0pt}

\begin{figure}
\centering
\includegraphics[scale=0.38]{"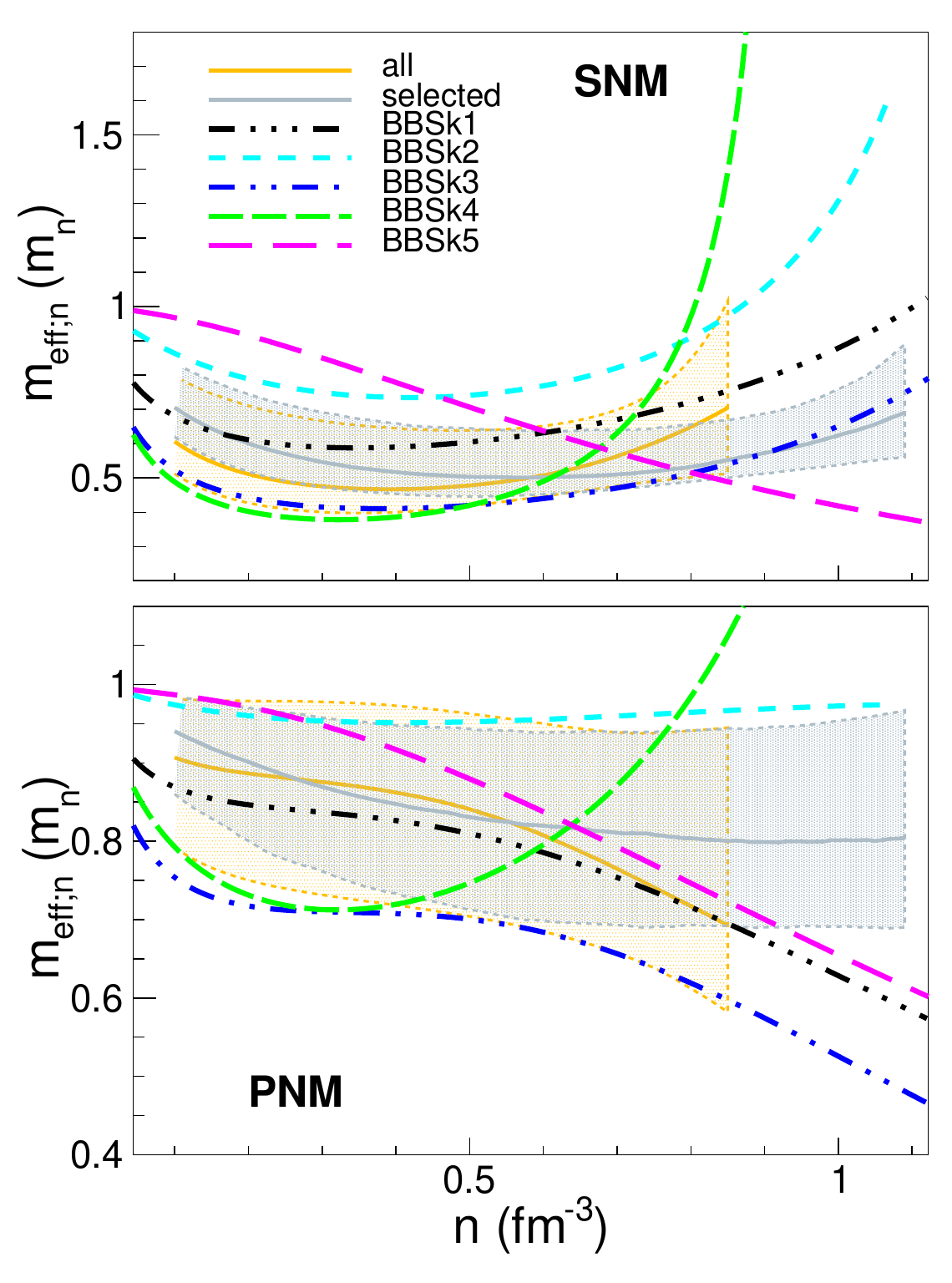"}
 \caption{Neutron effective mass ($m_{\mathrm{eff;}n}$) in SNM (top) and PNM (bottom) in units of bare neutron mass ($m_n$) as a function of density.
  Medians and upper and lower quantiles at 90\% CI of the two sets of models in Sec.~\ref{sec:NM} are depicted with solid and short dashed curves, respectively.
  The other curves correspond to BBSk1 - BBSk5 forces, see Paper I.
  }  
 \label{Fig:meffn}
\end{figure}

\begin{figure}
\centering
\includegraphics[scale=0.38]{"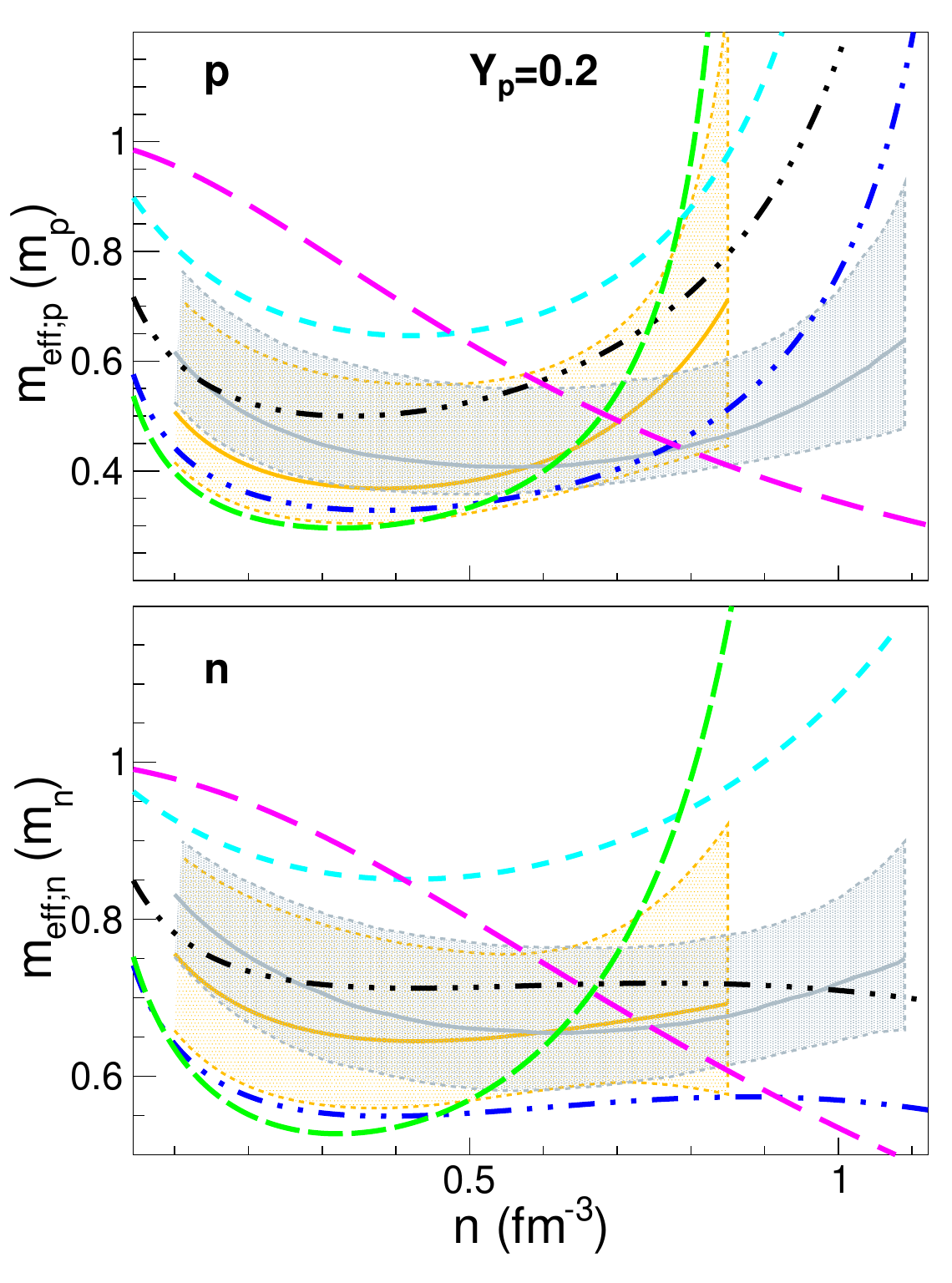"}
\caption{Neutron ($m_{\mathrm{eff;}n}$) and proton ($m_{\mathrm{eff;}p}$) effective masses
  in units of bare neutron ($m_n$) and proton ($m_p$) masses, respectively
  as functions of density in NM matter with $Y_p=0.2$.
  For the legend, see Fig.~\ref{Fig:meffn}.
  }  
 \label{Fig:meff_Yp=0.2}
\end{figure}

\begin{figure}
\centering
\includegraphics[scale=0.38]{"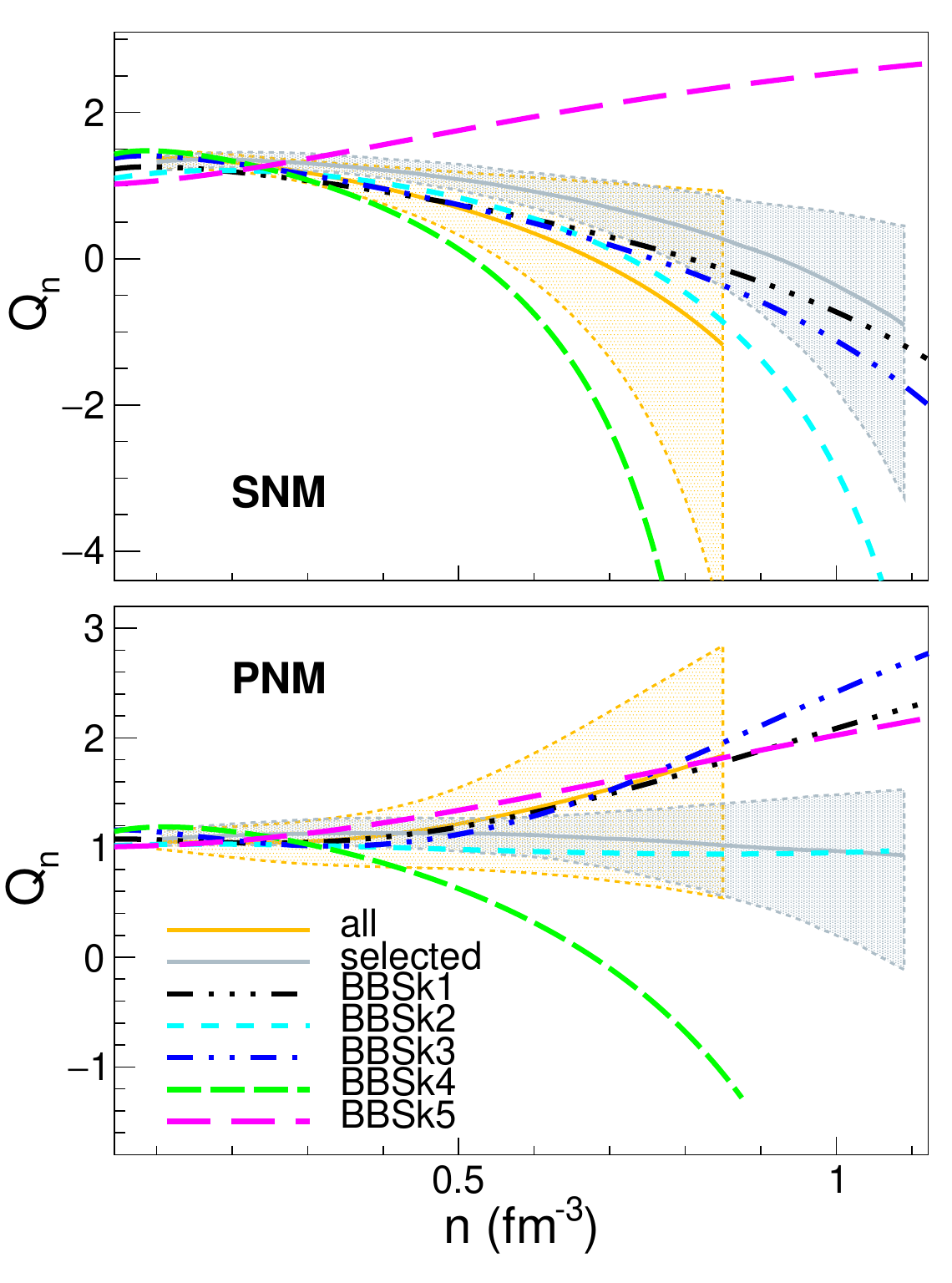"}
 \caption{The same as in Fig.~\ref{Fig:meffn} but for $Q_n$, see Eq.~\eqref{eq:Qi}.}
 \label{Fig:Qn}
\end{figure}

In this work, finite-temperature effects are investigated for two sets of effective interactions as well as the five effective interactions for which general purpose EOS tables were previously built in Paper I.

The first set (S1) corresponds to the ensemble of effective interactions from the run~1 in \citep{Beznogov_PRC_2024}.
These interactions have been generated within a Markov Chain Monte Carlo (MCMC) procedure that constrained both nuclear and NS matter. 
For NM we constrained:
i) the behavior of saturated SNM via the values of the saturation density ($\sat{n}$) and the energy per particle ($\sat{E}$) and compression modulus ($\sat{K}$) at $\sat{n}$,
ii) the symmetry energy ($\sym{J}$) at $\sat{n}$,
iii) the energy per neutron ($E/A$) in PNM at $n=0.08,~0.12$ and $0.16~\mathrm{fm}^{-3}$,
for which $\chi$EFT data by \cite{Drischler_PRC_2021} have been employed;
iv) for $n \leq n_l$, $0 \leq \meffnSNM /m_n \leq 1$ and  $0 \leq \meffnPNM /m_n \leq 1$,
v) for $n \leq n_l$, $v_{F;n}^\mathrm{SNM}/c \leq 1$ and $v_{F;n}^\mathrm{PNM}/c \leq 1$~\citep{Duan_PRC_2023}, where $v_{F;i}=\hbar k_{F;i}/m_{\mathrm{eff;}i}$ stands for the Fermi velocity of the $i$-particle.
We required NS EOSs to be
vi) causal up to the density that corresponds to the central density of the maximum mass configuration ($n_{\mathrm{c;TOV}}$), 
vii) thermodynamically stable,
viii) produce maximum gravitational masses ($M_{\mathrm{G;TOV}}$) in excess of $2~\Msun$.
For the upper boundary of the density domain where conditions iv) and v) are imposed, the value of $n_l=0.8~\mathrm{fm}^{-3}$ was chosen.
This value represented a compromise between the extension of the validity domain of the interaction and the computational efficiency.

The second set (S2) corresponds to the effective interactions in the set above that:
a) satisfy conditions iv) and v) for both neutrons and protons at arbitrary proton fractions $0.0 \leq Y_p \leq 0.5$ and for $n \leq n_{\mathrm{c;TOV}}$,
b) satisfy condition vi) for arbitrary proton fractions $0.0 \leq Y_p \leq 0.5$ and $n \leq n_{\mathrm{c;TOV}}$.
The number of interactions in S2 represents $\approx 0.7\%$ of the number of interactions in S1.
Even if no constraint was imposed on the thermodynamic stability of NM with arbitrary isospin asymmetry, it turned out that this is automatically achieved for all interactions.

The interactions BBSk2, BBSk3, BBSk4, BBSk5 belong to S1 and feature extreme behaviors of effective masses and thermal pressure as functions of density. BBSk2 and BBSk3 lie close to the 0.98 upper and 0.02 lower quantile of $\eff{m}(n)$; BBSk4 and BBSk5 provide large negative and positive thermal pressures. BBSk1 belongs to run~2 in \citep{Beznogov_PRC_2024}, which, in addition to the constraints accounted for in run~1, also implements constraints on neutron effective mass in PNM and nucleon effective mass in SNM up to $n=0.16~\mathrm{fm}^{-3}$; for BBSk1, $\eff{m}^\mathrm{SNM}(n)$ corresponds to the ``median'' in that run.
For the values of NM parameters, see Table 2 in Paper I.

The properties of NM EOSs in the two sets are reported in Table~\ref{tab:Prop} in terms of medians and lower and upper quantiles at 90\% confidence interval (CI).
The extra constraints imposed in S2 filter out the models with the softest increase of the energy per particle ($E/A$) as a function of $n$ in SNM as well as those with the steepest increase of
$E/A(n)$ in PNM.
As a result, the symmetry energy ($\sym{E}$) is much reduced for densities in excess of $3 \sat{n}$. While most of the models in S1 manifest a steep increase with density, see Fig.~1 in \citep{Beznogov_PRC_2024}, for S2 the median only slightly increases with density (not shown). 
This means that almost half of the number of models in S2 have a decreasing $\sym{E}(n)$ at high densities.
The exclusion of models with soft EOSs for SNM is the consequence of the upper limit imposed on $\eff{m}$ at densities higher than $n_l$, see below.
The exclusion of models with stiff EOSs for PNM stems from the condition on PNM, which is stiffer than NS matter, to be causal up to densities higher than $n_l$ and results in slightly lower values of the maximum gravitational mass of NSs, see Sect.~\ref{sec:NS}.
Table~\ref{tab:Prop} shows that the extra constraints also entail an increase by 18\% of the value of the effective neutron mass in SNM at $0.16~\mathrm{fm}^{-3}$. 

Insight into the density dependence of the neutron effective mass in SNM and PNM is provided in Fig.~\ref{Fig:meffn}. Notice that, modulo the neutron-proton mass split, in SNM, $m_{\mathrm{eff;}n}^{\mathrm{SNM}}(n)=m_{\mathrm{eff;}p}^{\mathrm{SNM}}(n)$.
Here as well as in Figs.~ 2 to 11, the hatched areas show the 90\% CI domains corresponding to S1 (yellow) and S2 (grey), respectively.
It comes out that upon imposing the extra constraints of S2, the interactions that provide for $\meffnSNM$ low values over $n/\sat{n} \lesssim 4$ and high values over the complementary domain of density are suppressed. In other words, while the original set of interactions (S1) favors a rather pronounced U-shaped behavior of $\meffnSNM(n)$, the more restricted set (S2) features a weaker density dependence. Still, the median and both quantiles in S2 have U-shapes.
The extra constraints slightly reduce the dispersion bands of both $\meffnSNM(n)$ and $\meffnPNM(n)$.
For $n/\sat{n} \lesssim 4$, $\meffnPNM$ decreases with $n$ for both sets; at higher densities, $\meffnPNM$ in S1 (S2) continues to decrease (becomes density independent).
The extra constraints of S2 suppress two classes of interactions: those that provide low values of $\meffnPNM$ for $n/\sat{n} \lesssim 2$ and those with a steep decrease of $\meffnPNM$ at $n/\sat{n} \gtrsim 3$.
Here and in Figs.~2--11, we plot the quantiles and the bands up to the density where at least one model in a set violates the S2 conditions.

The density dependence of neutron ($m_{\mathrm{eff;}n}$) and proton ($m_{\mathrm{eff;}p}$) effective masses in neutron-rich matter is considered in Fig.~\ref{Fig:meff_Yp=0.2}.
For the two sets, the medians and the upper quantiles of $m_{\mathrm{eff;}n}$ and $m_{\mathrm{eff;}p}$ as well as the lower quantiles
of $m_{\mathrm{eff;}p}$ behave similarly to their $\meffnSNM(n)$-counterparts and are U-shaped.
The lower quantile of $m_{\mathrm{eff;}n}$ in S1 manifests two extrema, suggesting that many of the models that provide low values for this quantity do the same. The extra constraints in S2 discard all these models. 
The dispersions of $m_{\mathrm{eff;}n}$ and $m_{\mathrm{eff;}p}$ are not reduced much upon imposing the extra constraints.
$m_{\mathrm{eff;}p}$-curves manifest a stronger density dependence than $m_{\mathrm{eff;}n}$-curves.

For both sets of interactions, the dispersions of all considered $\eff{m}$ is significant even at $n/\sat{n}<1$.
Complementary information about the diversity of behaviors of $\eff{m}(n)$ accommodated by the MCMC procedure is offered by the BBSk1-BBSk5 interactions.
For BBSk2, $\meffnSNM$ along with $m_{\mathrm{eff;}n}$ and $m_{\mathrm{eff;}p}$ in neutron-rich matter are U-shaped; $\meffnPNM$ has practically no density dependence.
For BBSk3, $\meffnPNM$ has an inflection point at $n/\sat{n} \approx 2$ and a steep decrease with density for $n/\sat{n}>4$;
its $\meffnSNM(n)$ and $m_{\mathrm{eff;}p}(n)$ in neutron-rich matter are U-shaped;
its $m_{\mathrm{eff;}n}$ in neutron-rich matter features two extrema and a meager variation over $1 \lesssim n/\sat{n} \lesssim 6$.
For BBSk4, the density dependence of all considered effective masses is strong and of U-shape type.
For BBSk5, all the effective masses considered here decrease steeply with density.
Neither BBSk4 nor BBSk5 is compatible with the constraints in S2.
$\eff{m}(n)$ in BBSk1 is similar to that in BBSk3, but the values are different.

Extra insight into the behavior of the effective masses as well as thermal response, see Sec.~\ref{sec:FiniteT}, can be obtained following the density dependence of 
\be
Q_i=1- \frac32 \frac{n}{m_{\mathrm{eff};i}} \frac{\partial m_{\mathrm{eff};i}}{\partial n}~.
\label{eq:Qi}
\ee
Fig.~\ref{Fig:Qn} depicts the case of $Q_n$ for the situations considered in Fig.~\ref{Fig:meffn}. As the majority of the interactions in S1 has an U-shaped behavior of $\meffnSNM(n)$, for half of the models in this set $Q_n^{\mathrm{SNM}}<0$ for $n/\sat{n} \gtrsim 4.4$. The filtering out by the conditions in S2 of the interactions that provide a steep increase of $\meffnSNM(n)$ at high density results in a more moderate decrease of $Q_n^{\mathrm{SNM}}$ with density. Indeed, the median of the models in S2 reaches zero only at $n \approx 0.9~\mathrm{fm}^{-3}$.
The dominant decrease of $\meffnPNM(n)$ with density for the models in S1 makes that for this set even the lower quantile of $Q_n^{\mathrm{PNM}}$ stays positive.
The presence in S1 of models with exotic behaviors is signaled by BBSk5 and BBSk4:
for BBSk5 (BBSk4), $Q_n^{\mathrm{SNM}}$ and $Q_n^{\mathrm{PNM}}$ increase (decrease) with $n$.
For BBSk4, $Q_n^{\mathrm{SNM}}$ and $Q_n^{\mathrm{PNM}}$ become negative at $n \gtrsim 0.5~\mathrm{fm}^{-3}$ and $n \gtrsim 0.7~\mathrm{fm}^{-3}$, respectively.

\section{Neutron star matter}
\label{sec:NS}

\begin{figure}
\centering
\includegraphics[scale=0.38]{"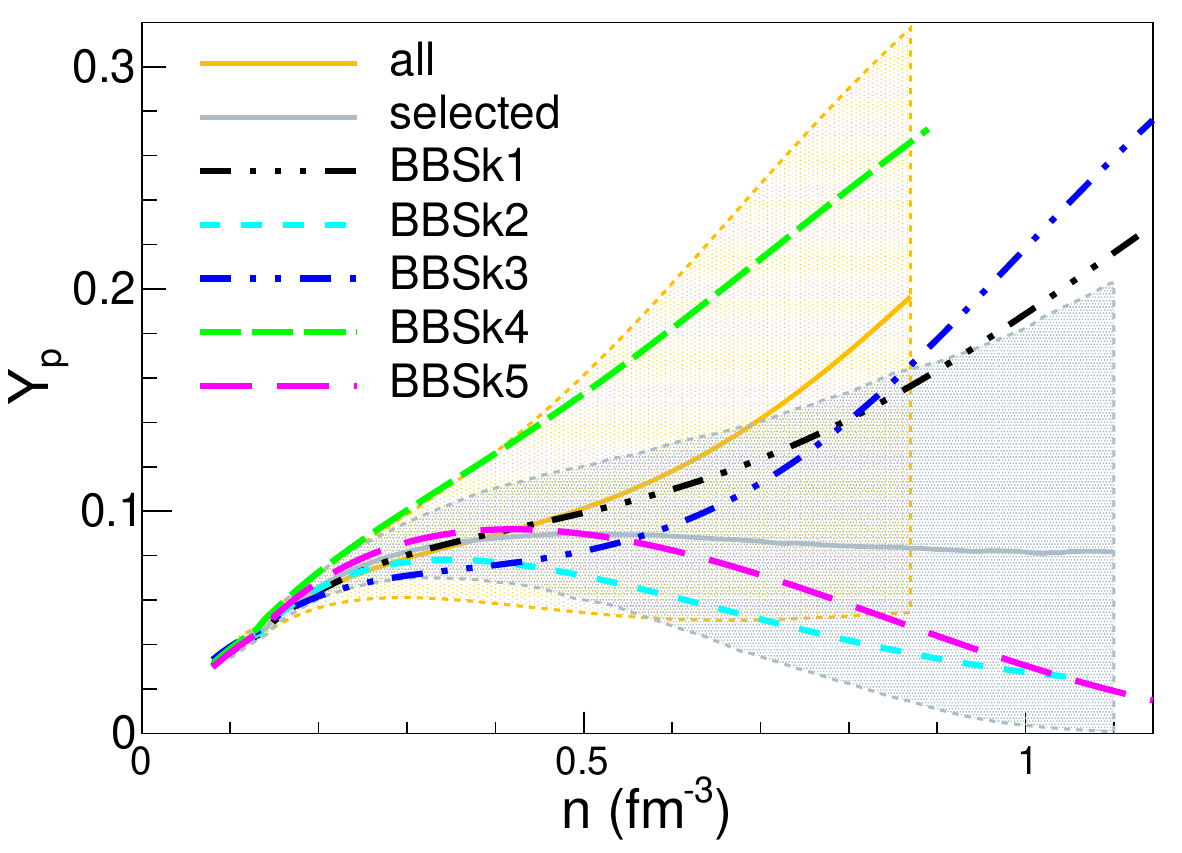"}
\caption{Proton fraction ($Y_p$) as a function of density in NS matter.
  }
 \label{Fig:Yp_NS}
\end{figure}

The values of key NS parameters corresponding to the two sets of interactions are reported in Table~\ref{tab:Prop}.
As already discussed in Sec.~\ref{sec:NM}, the requirement for PNM to be causal up to $n_{\mathrm{c;TOV}}$ softens the EOSs, which results in slightly lower values of $M_{\mathrm{G;TOV}}$ and $M_{\mathrm{B;TOV}}$ as well as higher values for the central density ($n_{\mathrm{c;TOV}}$), energy density ($\rho_{\mathrm{c;TOV}}$) and pressure ($P_{\mathrm{c;TOV}}$) of the most massive configurations. 
The speed of sound at $n_{\mathrm{c;TOV}}$ increases as well.
The medians of the $1.4\,\Msun$ and $2\,\Msun$ NSs' radii and tidal deformabilities are marginally affected by the extra constraints.
This suggests that the softening of PNM EOSs is compensated by the stiffening entailed by more neutron-rich cores due to much lower values of the symmetry energy.

The composition of NS matter is addressed in Fig.~\ref{Fig:Yp_NS} in terms of the proton fraction, $Y_p$, as a function of density.
The interactions in S1 allow for extremely different compositions. For most of the models, $Y_p$ increases with $n$.
Models like BBSk2 and BBSk5, which show the opposite trend, exist as well.
The strong reduction of the symmetry energy for $n/\sat{n} \gtrsim 3$  in S2 makes that the interactions with decreasing $Y_p(n)$ are slightly more numerous than those with increasing $Y_p(n)$.

\section{Thermal behavior}
\label{sec:FiniteT}

\begin{figure}
\includegraphics[scale=0.38]{"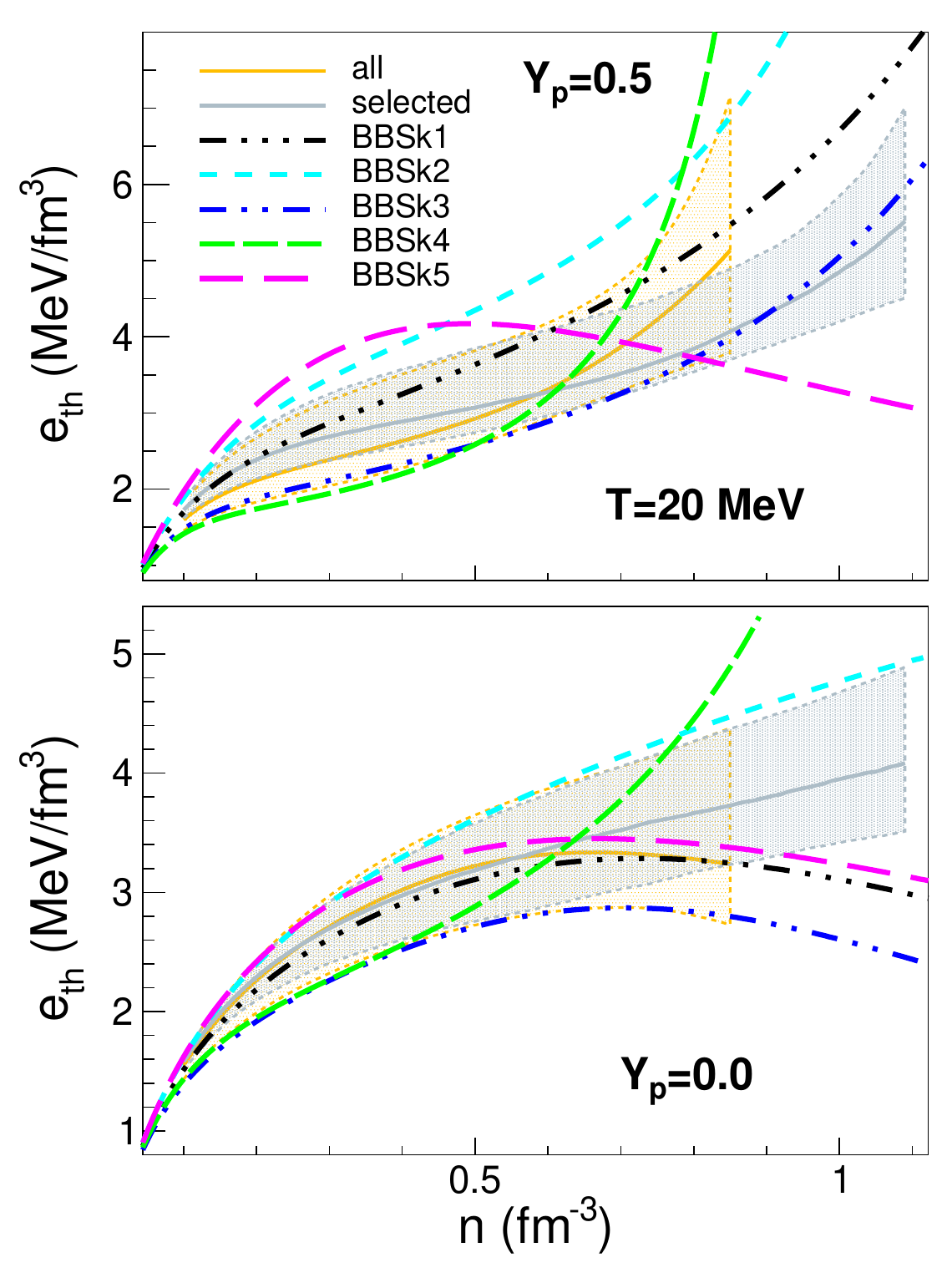"}
\centering
 \caption{$\tth{e}(n)$ in PNM and SNM matter at $T=20~\mathrm{MeV}$.
   }
 \label{Fig:eth_NM}
\end{figure}

\begin{figure}
\centering
\includegraphics[scale=0.38]{"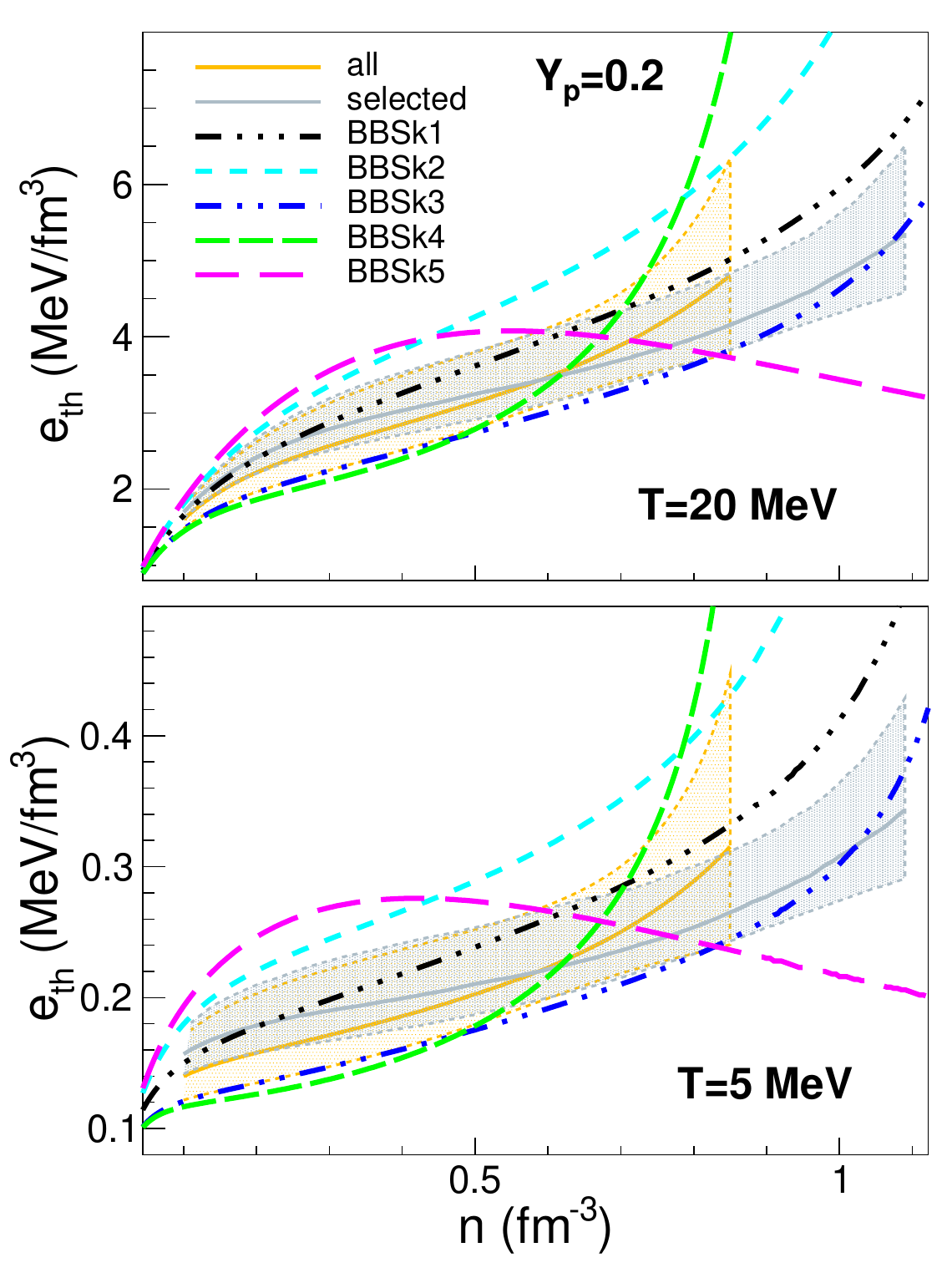"}
 \caption{The same as in Fig.~\ref{Fig:eth_Yp=0.2} but for NM matter with $Y_p=0.2$ for $T=5$ and 20 MeV.}
 \label{Fig:eth_Yp=0.2}
\end{figure}

\begin{figure}
\centering
\includegraphics[scale=0.38]{"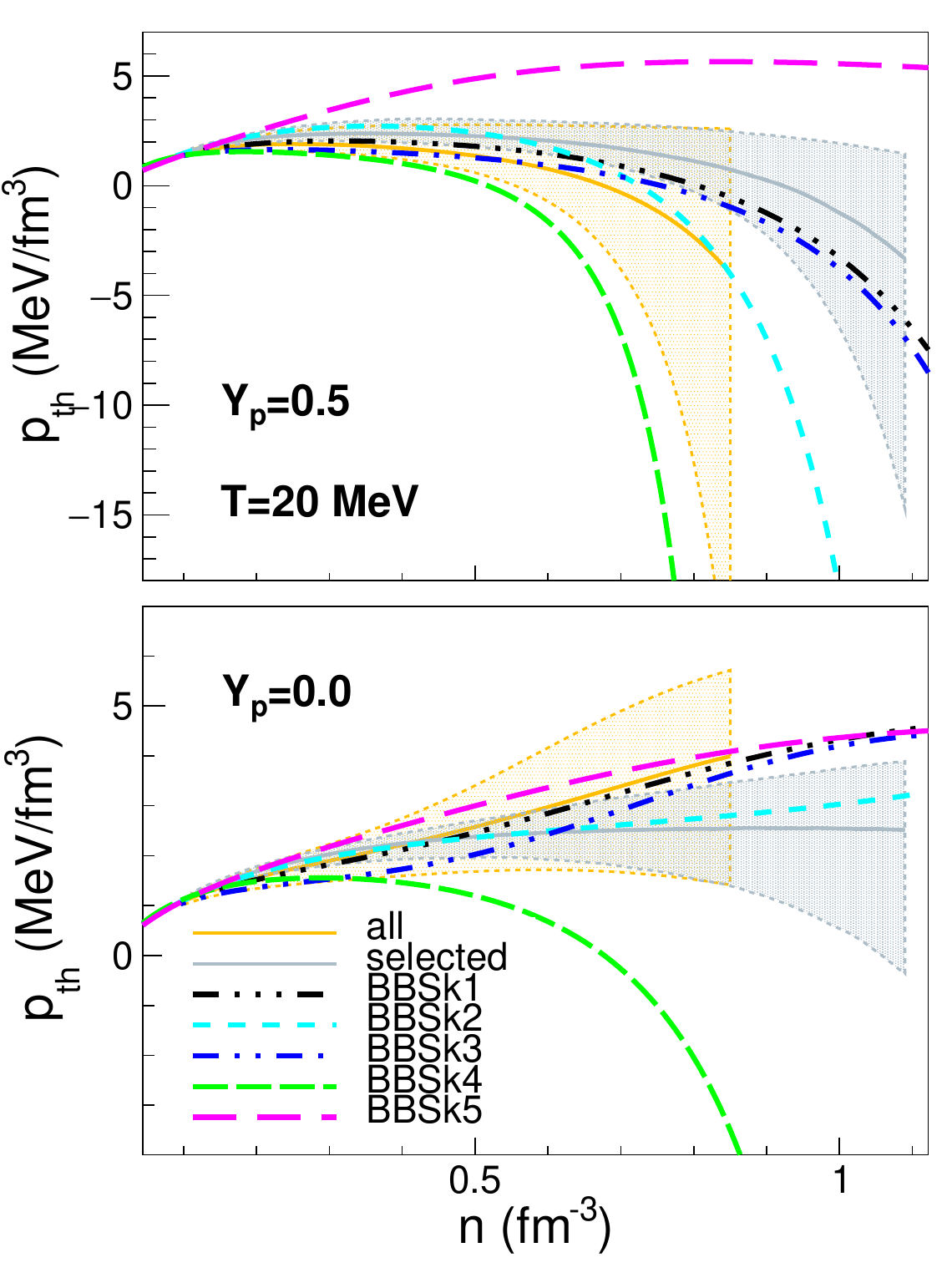"}
 \caption{$\tth{p}(n)$ in PNM and SNM matter at $T=20~\mathrm{MeV}$.
  }
 \label{Fig:pth_NM}
\end{figure}

\begin{figure}
\centering
\includegraphics[scale=0.38]{"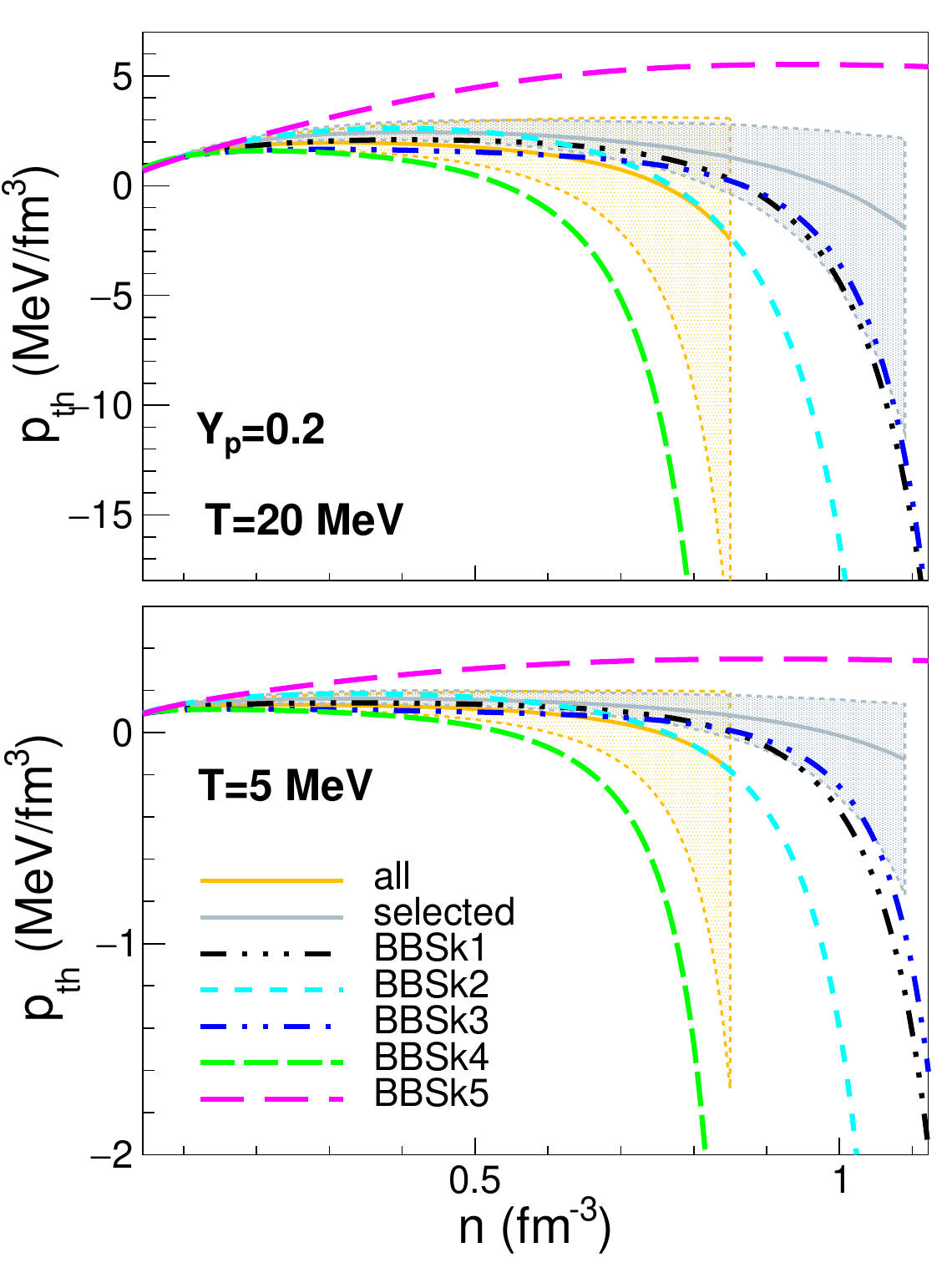"}
 \caption{ The same as in Fig.~\ref{Fig:pth_NM} but for NM matter with $Y_p=0.2$ for $T=5$ and 20 MeV.}
 \label{Fig:pth_Yp=0.2}
\end{figure}

\begin{figure}
\centering
\includegraphics[scale=0.38]{"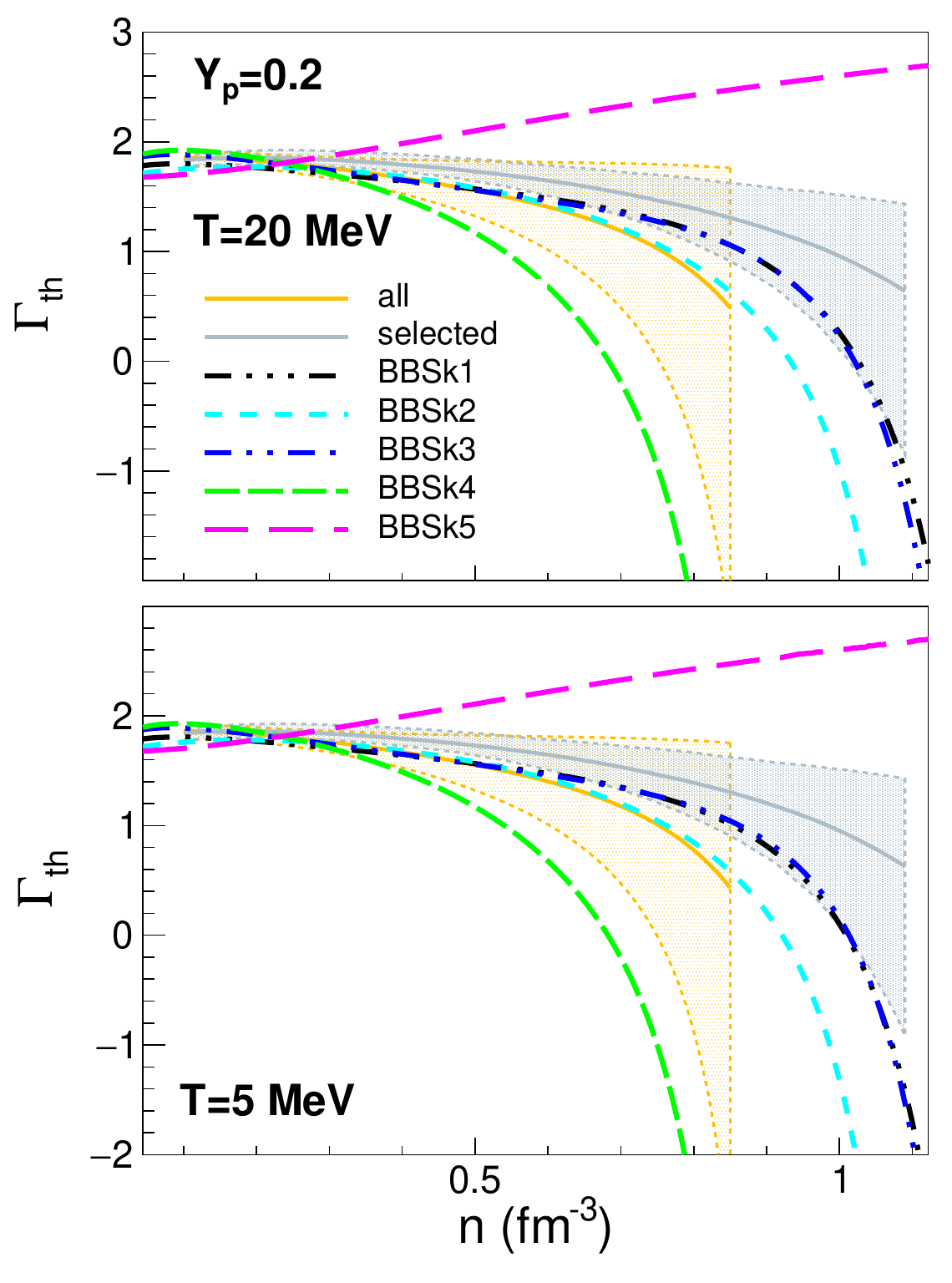"}
 \caption{$\tth{\Gamma}(n)$ in NM matter with $Y_p=0.2$ for $T=5$ and 20 MeV.
  }
 \label{Fig:Gth_Yp=0.2}
\end{figure}

\begin{figure}
\centering
\includegraphics[scale=0.38]{"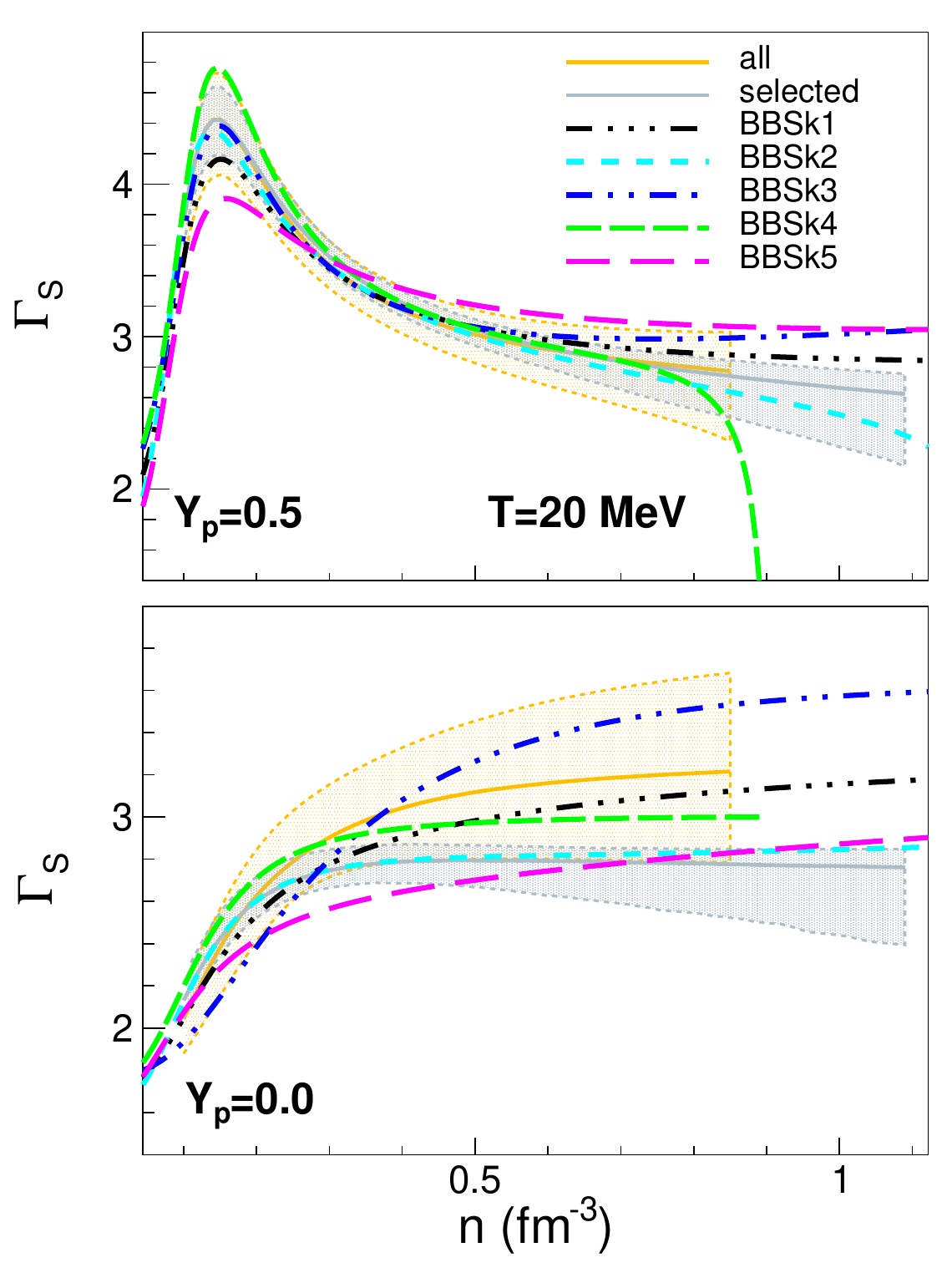"}
 \caption{$\Gamma_S(n)$ in PNM and SNM matter at $T=20~\mathrm{MeV}$.
 }
 \label{Fig:GS_NM}
\end{figure}

\begin{figure}
\centering
\includegraphics[scale=0.38]{"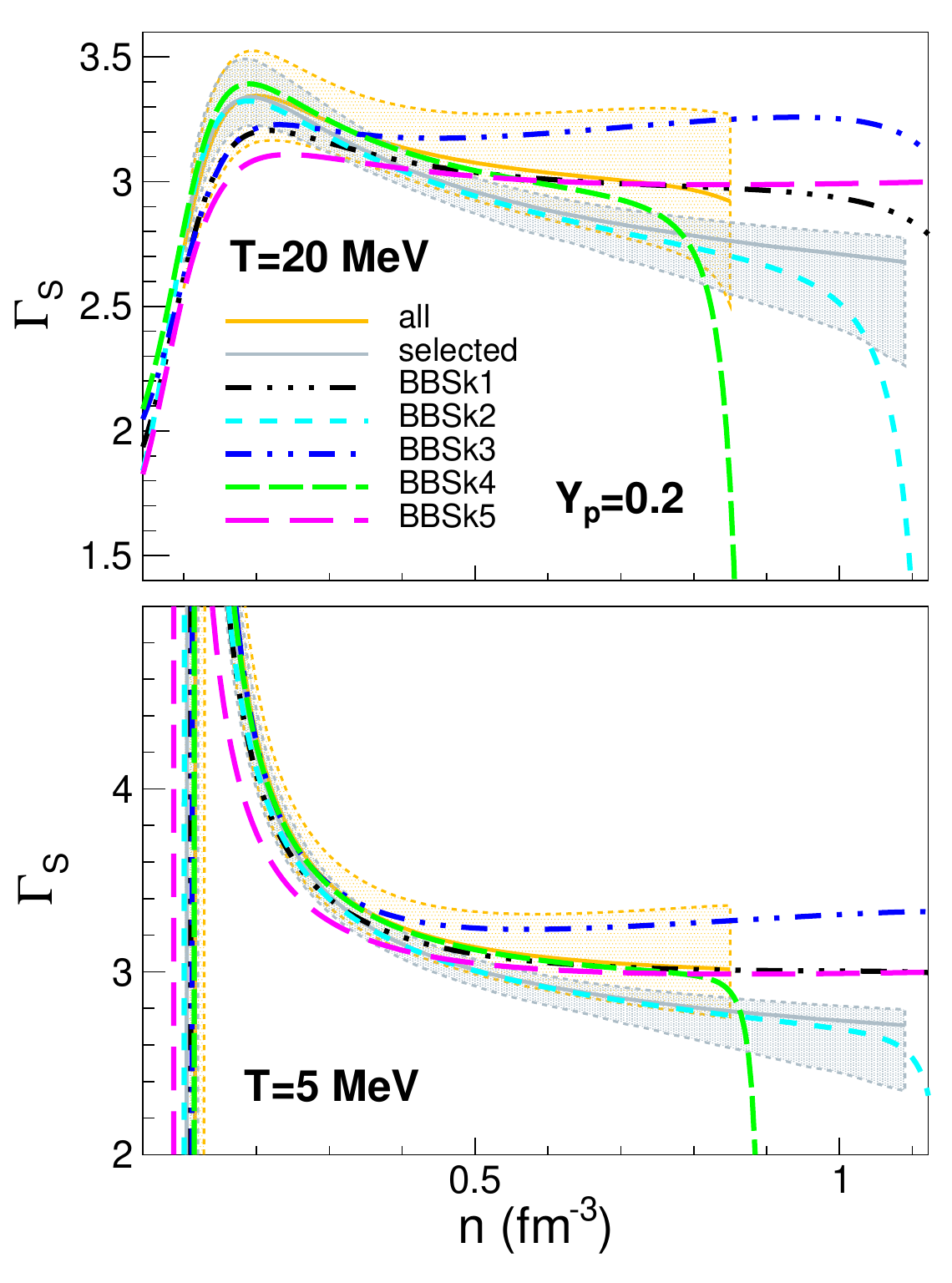"}
 \caption{The same as in Fig.~\ref{Fig:GS_Yp=0.2} but for NM matter with $Y_p=0.2$ for $T=5$ and 20 MeV.
  }
 \label{Fig:GS_Yp=0.2}
\end{figure}

We now turn to the finite-temperature behavior of NM. 
Following \cite{Constantinou_PRC_2014,Constantinou_PRC_2015,Raduta_EPJA_2021,Raduta_EPJA_2022}, the
density dependence of $\tth{e}$, $\tth{p}$, thermal ($\tth{\Gamma}$) and adiabatic ($\Gamma_S$) indexes will be investigated over wide domains of temperature and proton fraction. The same sets of interactions and specific forces as in the previous sections will be used. 

Figs.~\ref{Fig:eth_NM} and \ref{Fig:eth_Yp=0.2} address the density dependence of $\tth{e}$, see Eq.~\eqref{eq:eth}, in PNM and SNM with $T=20~\mathrm{MeV}$ and NM with $Y_p=0.2$ at $T=5,~20~\mathrm{MeV}$, respectively.
For all considered $(Y_p,T)$-sets and $n/\sat{n} \lesssim 2$, $\tth{e}(n)$ increases.
At higher densities, several behaviors exist.
For NM with $Y_p=0.2$ and SNM, most of the interactions predict that $\tth{e}(n)$ increases. 
An exception from this trend is offered by BBSk5, which features a steady decrease.
In PNM, the rise and fall behavior is common for most of the interactions belonging to S1, see the median as well as the predictions of BBSk3 and BBSk5. 
Models predicting that $\tth{e}(n)$ increases exist as well, see the curves corresponding to BBSk2 and BBSk4 and the upper quantile.
BBSk1, which corresponds to the median behavior of a run which imposes additional constraints on $\eff{m}$, also leads to a rise-and-fall behavior.
Upon imposing the extra conditions in S2, practically all models show an increasing $\tth{e}(n)$.
Remarkably, in all cases, the ordering of $\tth{e}$ at fixed $n$ replicates the one of $\eff{m}$.
The strongest density dependence is obtained for BBSk4, which also provides the strongest density dependence of $\eff{m}$. 
The relative widths of the uncertainty bands stay constant over the considered $T$-range.
For S2 and 90\% CI, its values at $n=0.8~\mathrm{fm}^{-3}$ are:
$0.31$ ($0.30$), $0.26$ ($0.25$), $0.31$ ($0.30$)
for $Y_p=0$, $0.2$ and $0.5$ and $T=5~\mathrm{MeV}$ ($T=20~\mathrm{MeV}$).

The density dependence of $\tth{p}$, see Eq.~\eqref{eq:pth}, in PNM and SNM with $T=20~\mathrm{MeV}$ and NM with $Y_p=0.2$ at $T=5,~20~\mathrm{MeV}$ is demonstrated in Figs.~\ref{Fig:pth_NM} and \ref{Fig:pth_Yp=0.2}, respectively.
Because of the U-shaped dependence of $\eff{m}(n)$, see Figs.~\ref{Fig:meffn} and \ref{Fig:meff_Yp=0.2}, at high densities many interactions predict a strong decrease of $\tth{p}$ up to negative values. 
$\tth{p}<0$ occurs more frequently in SNM than in PNM; $\tth{p}(n)$ of NM with $Y_p=0.2$ resembles more the behavior in SNM than the one in PNM. 
This means that the behavior of $\tth{p}$ is not determined by the most abundant species but by the one with the strongest density dependence. 
The most extreme behaviors correspond to BBSk4 and BBSk5.
For high densities and $Y_p \geq 0.2$, even the more restricted set, S2, features a significant number of models with $\tth{p}<0$.
The relative width of the uncertainty band increases with the proton fraction and stays constant with the temperature.
For S2 and 90\% CI, its values at $n=0.8~\mathrm{fm}^{-3}$ are:
$0.72$ ($0.71$), $1.64$ ($1.62$), $2.56$ ($2.56$)
for $Y_p=0$, $0.2$ and $0.5$ and $T=5~\mathrm{MeV}$ ($T=20~\mathrm{MeV}$).

The thermal index,
\be
\tth{\Gamma}=1+\frac{\tth{P}}{\tth{e}}
\label{eq:Gammath}
\ee
is commonly used to gauge the departure from the ideal gas behavior as well as to supplement cold EOSs with thermal contributions~\citep{Bauswein_PRD_2010,Hotokezaka_PRD_2013,Endrizzi_PRD_2018,Camelio2019,Weih_PRL_2020}.

Fig.~\ref{Fig:Gth_Yp=0.2} investigates the behavior of $\tth{\Gamma}(n)$ in NM with $Y_p=0.2$ at $T=5,~20~\mathrm{MeV}$.
In the limit of low densities, where interactions are negligible, $\tth{\Gamma} \to 5/3$, which is the ideal gas limit. 
For $n \lesssim 0.2~\mathrm{fm}^{-3}$ the dispersion among models is very low and the density dependence is weak. Still, the curves corresponding to BBSk4 and BBSk5 clearly indicate that BBSk4 (BBSk5) provides a $\tth{\Gamma}$ that decreases (increases) with $n$.
As a matter of fact, these trends are preserved at higher densities.
For $n/\sat{n} \gtrsim 2$, our models manifest an important density and interaction dependence. At $n=0.7~\mathrm{fm}^{-3}$ and $T=5~\mathrm{MeV}$ the most extreme values are $-0.2$ (for BBSk4) and $2.3$ (for BBSk5). 
Similar values are obtained for $T=20~\mathrm{MeV}$, which means that the $T$-dependence is negligible. 
Eqs.~\eqref{eq:eth} and \eqref{eq:pth} show that for the particular cases of SNM and PNM, $\tth{\Gamma}$ does not depend on $T$.
Indeed, in these two situations, $\tth{\Gamma}=1+2/3 Q_n$. 
The median of models in S1 (S2) features a pronounced (moderate) decrease with density, which stems from the strongly (moderately) U-shaped behavior of $m_{\mathrm{eff;}n}$ and $m_{\mathrm{eff;}p}$ for most of the models in this set.
The negative values of $\tth{\Gamma}$ are due to the negative values of $\tth{p}$.
Considering that, in suprasaturated stellar matter, the dominant contribution to pressure comes from nucleons, $\tth{\Gamma}$ stays negative even after taking electrons into account.
As expected based on the behaviors of $\tth{e}$ and $\tth{p}$, $\tth{\Gamma}$ of NM with $Y_p=0.2$ resembles $\tth{\Gamma}$ of SNM (not shown) while the behavior of $\tth{\Gamma}$ of PNM (not shown) is dissimilar.

The adiabatic index,
\be
\Gamma_S=\left.\frac{\partial \ln P}{\partial \ln n} \right|_S,
\label{eq:GammaS}
\ee
measures the EOSs' stiffness in isentropic processes. Its behavior in PNM and SNM with $T=20~\mathrm{MeV}$ and NM with $Y_p=0.2$ at $T=5,~20~\mathrm{MeV}$ is investigated in Figs.~\ref{Fig:GS_NM} and \ref{Fig:GS_Yp=0.2}, respectively.
For temperatures lower than an interaction- and $Y_p$-dependent value, subsaturated SNM and NM with moderate isospin asymmetries feature a liquid-gas phase instability~\citep{Ducoin_NPA_2006}. This instability manifests as a back-bending in all $Y_i(X_i)_{Y_j,~j\neq i}$ curves, where $X_i$ and $Y_i$ denote an extensive variable and its conjugated intensive variable, respectively~\citep{Ducoin_NPA_2006}.
The divergence of $\Gamma_S$ in NM with $Y_p=0.2$ and $T=5~\mathrm{MeV}$ is the outcome of the back-bending behaviors of $P(n)_{T}$.
Also, the pronounced peak of $\Gamma_S$ in SNM with $T=20~\mathrm{MeV}$ echos the vicinity of the critical temperature of the liquid-gas phase transition, whose typical values are $16~\mathrm{MeV} \lesssim T_C \lesssim 20~\mathrm{MeV}$.
These instabilities, however, are not relevant in stellar matter, which gets stabilized by clusterization, see Paper I.
For $n/\sat{n} \geq 2$, most models provide a smooth and weak variation of $\Gamma_S(n)$.
For BBSk2 and BBSk4, $\Gamma_S$ features a sudden drop at densities close to the validity limit of the models.
Upon imposing the extra conditions in S2, most of the models predict a slow decrease of $\Gamma_S(n)$. Also, the dispersion among the models is relatively low. 

\section{Stability of proto-neutron stars and remnants of binary neutron star mergers}
\label{sec:PNS}

\begin{figure}
\centering
\includegraphics[scale=0.38]{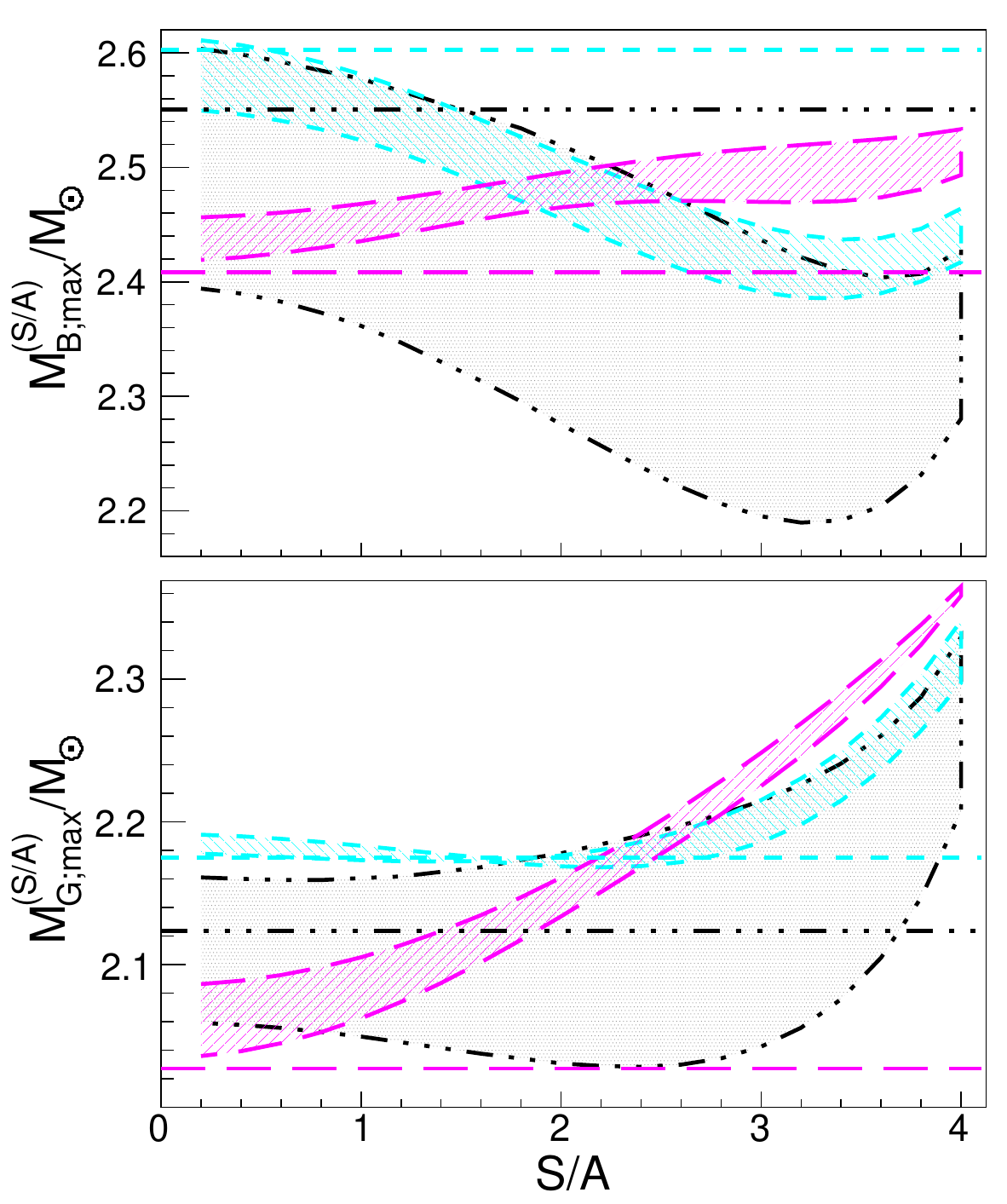}
 \caption{Maximum gravitational (bottom) and baryonic (top) masses versus entropy per baryon ($S/A$) for isentropic PNSs with constant profiles of $0.06 \leq Y_p \leq 0.3$, as predicted by BBSk1 (black), BBSk2 (cyan) and BBSk5 (magenta) forces.
 Horizontal lines mark the corresponding values, $M_{\mathrm{G;TOV}}$ and $M_{\mathrm{B;TOV}}$, for cold catalyzed NSs.
 }
 \label{Fig:Mmax}
\end{figure}

\begin{figure}
\centering
\includegraphics[scale=0.42]{"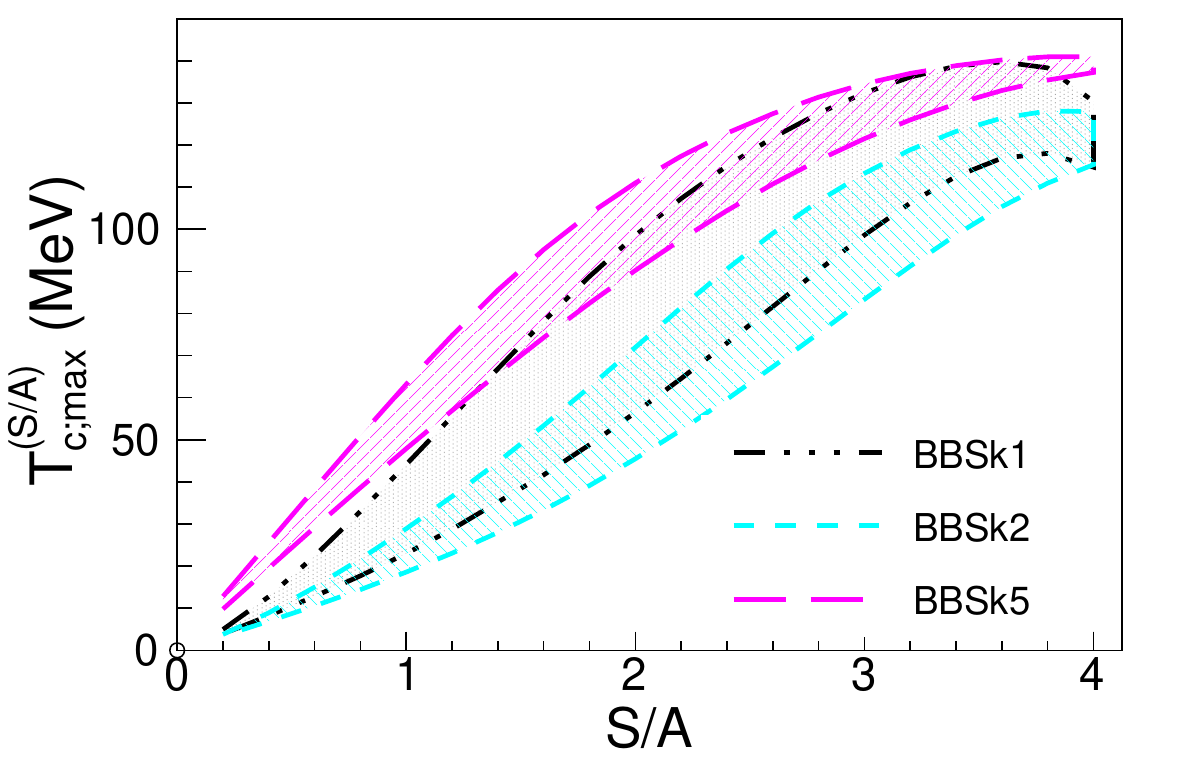"}
 \caption{Central temperature of the maximum mass configuration ($T_{\mathrm{c;max}}^{(S/A)}$) versus entropy per baryon ($S/A$) for the same cases as in Fig.~\ref{Fig:Mmax}. 
 }
 \label{Fig:Tncmax}
\end{figure}

The maximum gravitational mass of isentropic compact stars is relevant for BH formation in failed CCSNe.
Numerical simulations by \cite{Schneider_ApJ_2020} indicate that the onset of collapse coincides with the instance when PNS’s gravitational mass exceeds the maximum gravitational mass corresponding to its most common (throughout the volume) entropy value.

Assuming the absence of accretion and mass loss, the baryonic mass is conserved, making it a useful quantity in the context of analyzing the stability of PNSs and BNS mergers remnants. Obviously, if the baryonic mass of a PNS (or remnant) exceeds the maximum baryonic mass of the cold beta-equilibrated NS, the PNS (or remnant) is necessarily unstable with respect to collapse into a BH regardless of the mechanism that might temporarily stabilize it.

The relation between the maximum gravitational mass of isentropic stars, $M_{\mathrm{G;max}}^{(S/A)}$ and the maximum gravitational mass of cold-catalyzed configurations, $M_{\mathrm{G;TOV}}$,
was considered by \cite{Raduta_MNRAS_2020} and \cite{Wei_PRC_2021}, who employed a large number of phenomenological and microscopic models. 
Their results show that for models based on the covariant density functional theory of NM and for microscopic variational models, $M_{\mathrm{G;max}}^{(S/A)} > M_{\mathrm{G;TOV}}$ (for all values of $S/A$), while the relation gets inverted for models based on the microscopic Brueckner-Hartree-Fock theory.
$M_{\mathrm{G;max}}^{(S/A)} > M_{\mathrm{G;TOV}}$ (for all values of $S/A$) is also obtained for Skyrme-like interactions with $\eff{m}$ equal to the bare mass and $\sat{K}$ within the presently accepted range.
The situation of the maximum baryonic mass of hot stars is ambiguous.
According to \cite{Raduta_MNRAS_2020} models based on the covariant theory of NM lead to both $M_{\mathrm {B;max}}^{(S/A)}$ increasing and decreasing with the value of entropy per baryon.

Fig.~\ref{Fig:Mmax} illustrates $M_{\mathrm{G;max}}^{(S/A)}$ and $M_{\mathrm {B;max}}^{(S/A)}$ as functions of entropy per baryon ($S/A$), as predicted by the BBSk1, BBSk2 and BBSk5 effective interactions.
For simplicity, we also assume that hot compact objects have constant radial profiles of $Y_p$, and we vary these values between $0.06 \leq Y_p \leq 0.3$ (arbitrary) forming the bands demonstrated in the figure. 
It turns out that for BBSk1, both $M_{\mathrm{G;max}}^{(S/A)}$ and $M_{\mathrm {B;max}}^{(S/A)}$ show a strong dependence on $Y_p$, while for BBSk2 and BBSk5 the dependence is rather weak. 
This result is the obvious consequence of high (low) values of the symmetry energy in BBSk1 (BBSk2 and BBSk5), see Sec.~\ref{sec:NM}. 
The three interactions also manifest different evolution of $M_{\mathrm{G;max}}^{(S/A)}$ and $M_{\mathrm {B;max}}^{(S/A)}$ as functions of $S/A$. BBSk1 and BBSk2 predict that up to a certain value $M_{\mathrm{G;max}}^{(S/A)}$ ($M_{\mathrm {B;max}}^{(S/A)}$) stays constant (decreases) and then increases; for BBSk5, both $M_{\mathrm{G;max}}^{(S/A)}$ and $M_{\mathrm {B;max}}^{(S/A)}$ increase with $S/A$ although with different slopes.
For BBSk1, $M_{\mathrm{G;max}}^{(S/A)} \gtrless M_{\mathrm{G;TOV}}$ ($M_{\mathrm{B;max}}^{(S/A)} \gtrless M_{\mathrm{B;TOV}}$) depending on $Y_p$ ($Y_p$ and $S/A$).
For BBSk2, for all $Y_p$ and $S/A$, $M_{\mathrm{G;max}}^{(S/A)} \gtrsim M_{\mathrm{G;TOV}}$ and $M_{\mathrm{B;max}}^{(S/A)} \lesssim M_{\mathrm{B;TOV}}$, the latter being highly unusual. 
For BBSk5, for all $Y_p$ and $S/A$, $M_{\mathrm{G;max}}^{(S/A)} > M_{\mathrm{G;TOV}}$ and $M_{\mathrm{B;max}}^{(S/A)} > M_{\mathrm{B;TOV}}$.

In view of the criterion mentioned in the first paragraph of this section, one can conclude that:
i) in a failed CCSN, a PNS with $S/A \gtrsim 2.6$ built on BBSk1 will collapse earlier than a PNS built on BBSk5,
ii) a PNS with $S/A \lesssim 2.2$ built on BBSk5 will collapse earlier than a PNS built on BBSk2.

Now, let us consider the criterion based on the maximum baryonic masses.
Here, the difference between the considered EOSs is even more striking.  
For BBSk5 the situation is rather standard: thermal support stabilizes the star against collapse.
The hotter the star, the higher is the maximum supported baryonic mass.
For BBSk2 the situation is exactly the opposite: a cold beta-equilibrated star supports higher values of the maximum baryonic mass than hot stars.
This means that, in the absence of other stabilizing mechanisms, the PNS (or remnant) exceeding $M_{\mathrm{B;max}}^{(S/A)}$ will collapse promptly into a BH even if its mass is less than the maximum baryonic mass of a cold NS.
This outcome is a direct consequence of $p_\mathrm{th}$ being negative (in some $T - n - Y_p$ domain) for BBSk2 compared to $p_\mathrm{th}$ staying always positive for BBSk5, see Sec.~\ref{sec:FiniteT}. 
A similar effect was also observed in some Brueckner-Hartree-Fock models~\citep{Lu_PRC_2019}.

In agreement with \cite{Schneider_PRC_2019b,Yasin_PRL_2020,Schneider_ApJ_2020,Andersen_ApJ_2021,Fields_ApJL_2023}, these results confirm that the evolution of astrophysical phenomena that involve hot dense matter probes the nucleon effective mass and its density dependence.
However, the effective interactions used here have more sophisticated $\eff{m}(n)$ dependencies that render the link with $\eff{m}(\sat{n})$ not trivial. 
In addition, it is worth noting that for models with high symmetry energy at high densities (e.g., BBSk1) the stability of hot compact objects also depends on the $Y_p$ profile. 

Fig.~\ref{Fig:Tncmax} investigates the central temperature of the maximum mass configuration ($T_{\mathrm{c;max}}^{(S/A)}$) as a function of $S/A$ for the cases considered in Fig.~\ref{Fig:Mmax}. 
For $S/A \lesssim 3.5$, all forces predict that $T_{\mathrm{c;max}}^{(S/A)}$ increases with $S/A$; the spread of data increases with $S/A$, too.
For all effective interactions, the highest (lowest) value of $T_{\mathrm{c;max}}^{(S/A)}$ corresponds to the lowest (highest) value of $Y_p$.
The maximum spread for given $S/A$ corresponds to BBSk1 and is due to the high values of its symmetry energy for densities in excess of $3\sat{n}$.
In addition, high values of $\eff{m}$ at $n_{\mathrm{c;max}}$ are correlated with low values of $T_{\mathrm{c;max}}^{(S/A)}$.

\section{Conclusions}
\label{sec:concl}

A large number of ab initio constrained EOS models previously generated within a Bayesian analysis~\citep{Beznogov_PRC_2024} was used to investigate the thermal response of dense NM over domains of density, temperature, and proton fraction relevant for the evolution of astrophysical phenomena that involve hot compact objects.
These models rely on the Brussels parametrization of the Skyrme effective interaction, which presents two major avails.
First, it is flexible enough to allow for widely different behaviors in the suprasaturation regime, including a density dependence of the nucleonic effective mass in qualitative agreement with the predictions of microscopic calculations with three body forces~\citep{Baldo_PRC_2014,Burgio_PRC_2020,Drischler_PRC_2021}.
Second, the availability of analytical expressions for most thermodynamic and microscopic quantities makes it possible to assess the role that various NM parameters play at finite temperature.

The insufficient knowledge of the dense matter EOS, commented at length in the literature, gets translated also into thermal responses that differ much from one model to another.
In general, effective interactions that provide an U-shaped behavior of $\eff{m}(n)$ lead to EOSs that have much lower thermal pressure compared to those generated based on effective interactions that provide a monotonic decrease of $\eff{m}(n)$.
For densities in excess of several times $\sat{n}$, the former models feature negative thermal pressures.
$\tth{p}<0$ were previously obtained within $\chi$EFT calculations~\citep{Keller_PRL_2023}.

The stiffness of the EOS at finite temperatures impacts the stability of hot stars, with consequences for the fates of PNSs and BNS mergers.
Three particular interactions in our set, which manifest different behaviors of $\eff{m}(n)$, lead to different dependencies of the maximum gravitational and baryonic masses as a function of $S/A$.
In particular, a model whose $\eff{m}$ is decreasing with $n$, predicts that both $M_{\mathrm{G;max}}^{(S/A)}$ and $M_{\mathrm{B;max}}^{(S/A)}$ exceed their counterparts in cold beta-equilibrated NSs.
Two other models, which feature U-shaped behaviors of $\eff{m}(n)$, show that, depending on $S/A$, $Y_p$, and effective interaction, $M_{\mathrm{G;max}}^{(S/A)} \gtrless M_{\mathrm{G;TOV}}$, $M_{\mathrm{B;max}}^{(S/A)} \gtrless M_{\mathrm{B;TOV}}$ with $M_{\mathrm{B;max}}^{(S/A)} < M_{\mathrm{B;TOV}}$ being the most notable result.
The stability with respect to the collapse of models with high $\sym{E}$ at high densities also depends on the $Y_p$-profile. 

The use, in numerical simulations, of EOS tables based on Brussels-Skyrme interactions, e.g., those introduced in Paper I and available on \textsc{CompOSE}, will make the correlation between $\eff{m}$ and the evolution of astrophysical phenomena more difficult to establish than in \citep{Schneider_PRC_2019b,Yasin_PRL_2020,Schneider_ApJ_2020,Andersen_ApJ_2021,Fields_ApJL_2023}. Nonetheless, this is a necessary step toward a better understanding of the properties of dense matter, including the possible presence of non-nucleonic degrees of freedom.

The present work is a follow-up of the work by \cite{Raduta_PLB_2024}, who considered the thermal response of a set of EOSs that rely on the covariant density functional of nuclear matter and were generated within a Bayesian inference of the EOS of dense matter~\citep{Beznogov_PRC_2023}. 
In this way, we contribute to a better understanding of the role the EOS plays in the evolution of hot compact objects. 

\begin{acknowledgments}	
We acknowledge support from a grant from the Ministry of Education and Research, CNCS/CCCDI–UEFISCDI, Project No. PN-IV-P1-PCE-2023-0324;
partial support from Project No. PN 23 21 01 02 is also acknowledged.
The two authors have contributed equally to this work.
\end{acknowledgments}	

\bibliographystyle{aasjournal-hyperref}
\bibliography{BSk_T.bib}
\end{document}